\documentstyle[12pt,aaspp4,flushrt]{article}
 
\lefthead{Batalha et al.}
\righthead{Spectral Variability of GQ Lupi}

\begin{document}
 
\title{Variability of Southern T Tauri Stars I: The Continuum and the H$\beta$ Inverse PCygni Profile of GQ LUPI
}

\author{C. Batalha, D.F. Lopes and N.M. Batalha\altaffilmark{2}}
\authoraddr{Observat\'orio Nacional/CNPq, Rua General Jos\'e Cristino, 77, Rio de Janeiro, RJ 20920-400, Brazil}
\authoremail{celso@on.br, dalton@on.br \& natalie@starlight.arc.nasa.gov}

\begin{abstract}

We present time series spectrophotometric observations of GQ
Lupi, a typical representative of the YY Ori subgroup of T Tauri stars that 
show conspicuous inverse PCygni profiles. 
The data set consists of 32 exposures taken over 5 and 8 consecutive
nights of May and July 1998, respectively, and covers the spectral range
of 3100 \AA~ $< \lambda < 5100 $ \AA. The region redward and next to the
Balmer jump varies significantly on a night--to--night basis and 
the amplitude of such variability decreases sharply at $\lambda
>$ 4600 \AA~.  The Balmer continuum slope indicates that the spectral energy 
distribution is governed by a gas of temperature greater than that of the 
stellar photosphere. The variability of the Balmer continuum flux has the 
largest amplitude.  Flux increases in the B band are accompanied by 
concurrent flux increases in the U band. The contrary is not always
verified.  

The excess of continuum emission (veiling) for each exposure is computed 
throughout the spectral format. We find a tight anticorrelation 
between the veiling and the observed Balmer jump.  
We report the largest inverse PCygni profile ever observed at this resolution: the depth of the H$\beta$ absorption component is 
nearly twice the height of the peak emission.  Surprisingly, this
absorption vanishes a few nights later.
The time series of the redward absorption component behaves similarly to
the veiling time series:
the progressive weakening of the redward absorption is 
closely followed by a similar weakening of the excess
continuum emission. If the absorption component is de--veiled, the correlation strengthens.
Thus, large/small redward absorption = large/small veiling = small/large
 Balmer jump. 

We model the emitting region by a gas of uniform temperature and
density: each of the 32 exposures acting as a snapshot of such a region
for a given stellar rotational phase. We explore models of temperature
greater than 5000 K and number of hydrogen atoms (N$\_{H}$) larger than 
10$^{13}$ cm$^{-3}$, extending the gas spectral energy distribution up to 
the blackbody of a given temperature. The resulting models indicate that 
the gas densities and the respective temperatures are strongly 
anticorrelated. In addition, the model time series show that
the increase in the gas density is mirrored by an
increase of the projected emitting area (filling factor). 
Large/small gas densities and filling factors are 
characterized by high/low observed veiling.  As the accretion rate
fades from night-to-night, the observed veiling decreases, as 
does the gas density and the total projected emitting area.

\end{abstract}

\keywords{Accretion, accretion disks -- stars: T Tauri, pre Main Seq\"uence -- stars: YY Ori, individual (GQ Lupi)}

\section{Introduction}

According to current models of classical T Tauri stars (CTTS), the
steady migration of the gas in a circumstellar accretion disk is
suddenly halted at a few stellar radii by the stellar magnetic field
(Uchida \& Shibata 1984; Camenzind 1990; K\"{o}nigl 1991, Shu et al.
1994).  Material then free--falls onto the star, guided by stellar magnetic
field lines.  If an axisymmetric
bipolar configuration of the stellar magnetosphere is further invoked,
the circumstellar gas -- which is confined near the equatorial
orbital plane -- will end its collapsing trip on high latitude stellar
regions.  This scenario explains the small
projected rotational velocity ($v\sin i < 15$ km s$^{-1}$) found among CTTS. 
The field lines, 
crossing the circumstellar disk beyond the co-rotational radius,
meet the disk atmosphere rotating more slowly than the stellar surface
thereby braking the star and regulating the stellar angular
momentum during the early phases of stellar evolution 
(Bouvier et al. 1993; Edwards et al.
1993).  A natural consequence of such a paradigm is the occurrence of
absorption features in the red wings of the broad emission lines commonly
observed in CTTS. An inverse PCygni profile (IPC) will result
in those cases where a red--shifted absorption component
depletes the local continuum.

The inverse PCygni profiles were highlighted for the first time 
in pre--Main Sequence studies by Walker (1972). He detected IPC in 
the high Balmer lines of the star YY Orionis, dubbing a new sub--class of 
pre--Main Sequence objects after this star. The presence of the IPC in YY
Ori stars was interpreted as the result of quasi--radial accretion
operating during the very final stages of the protostellar hydrodynamic
evolution (Walker 1972; Appenzeller \& Wolf 1977).  This view was
discarded  after the discovery of a handful of CTTS which display lines 
with normal
PCygni profiles -- indicative of mass outflow -- as well as lines with 
inverse
PCygni profiles, indicating mass infall.
Occasionally, they would change from one to the other on time 
scales of days
(Krautter \& Bastian 1980).  Thus, a more complex mass motion
incorporating gas flow in nearly opposite directions had to be
developed.  Mundt (1984) suggests that material is falling back onto the
star after being ejected at velocities smaller than the escape
velocity.  Mass transfer from a secondary onto the YY Ori star is
also discussed in Grasdalen (1977). More recently, 
Edwards et al. (1994), analyzing high resolution spectra of
15 CTTS with a broad range of accretion rates, concludes that the 
large majority of these objects displays redshifted absorption or 
profile asymmetries indicative of mass infall. Thus, rather than being 
an exception to the rule, the YY Ori features are actually common 
properties of CTTS.

Bertout et al. (1982) reports on high resolution spectroscopy of two
bright YY Ori stars during 12 consecutive nights -- GQ Lupi being one of
them.  They find that the redshifted absorption component shows the
largest variability on a short time scale compared to the other
line components. These findings were reinforced by Aiad et al.
(1984) who conducted a temporal survey of a few YY Ori members. They
find that the time scale for the redshifted absorption variability
depends on the star and that these features are formed in a 
free--falling matter within a few stellar radii.  With
velocities larger than a few hundreds of kilometers per second, 
and subject to short term variability,
the IPC stands as a tracer of accretion mediated by magnetic field lines
(Aiad et al.  1984). Lehmann, Reipurth \& Brander (1995) observed EX
Lupi during a maximum in emission and a few months later when the star
had dimmed.  The IPC is only present during the
bright phase.  They interpret this outburst in terms of an increase
of mass accretion, with the consequent increase in the continuum flux
and the spectral veiling.  At maximum continuum emission, the
photospheric lines vanish and IPC's are clearly visible.

Line emission is also thought to be produced in the accretion funnel
(Calvet \& Hartmann 1992)  rather than in pure winds 
(Mundt 1984; Hartmann et
al. 1990). Hartmann, Hewett \& Calvet (1994) develop infall models with
a dipole configuration for the magnetic field geometry. They restore
most of the Balmer line features seen in CTTS such as the
lack of redshifted absorption in the H$\alpha$ line compared to the 
higher Balmer lines and the slightly blueshifted emission
peak of the Balmer lines.  However, Johns \& Basri (1995a) note that 
the nearly
symmetric profile and the full width at zero level of the average
H$\alpha$ emission of SU Aur cannot be reproduced in an accretion
funnel alone. Instead, they claim that the bulk H$\alpha$ emission is
produced in an almost stationary zone close to the star and that 
turbulence, not the velocity field, broadens the line. 

Time series analysis is extremely suitable for exploring the 
circumstellar
gas configuration for this class of
objects.  A single snapshot, at any spectral resolution, allows, at the
most, a short glance at an intricate gas flow. On the contrary, time
series analysis allows for the search of correlations among selected 
components of
line profiles, among several line groups,  and between line and
continuum fluxes of a given star (Bertout et al. 1982; Giampapa et al.
1993; Johns \& Basri 1995a,b; Petrov et al. 1996;
Gullbring et al. 1996; Johns \& Basri 1997; Hessman \& Guenther 1997;
Lago \& Gameiro 1998). SU Aur has been a common target for such 
analyses (Giampapa et
al. 1993; Johns \& Basri 1995; Petrov et al. 1996). Its H$\alpha$ line
profile has a blueshifted absorption feature with periodic behavior. 
A similar
feature is also found in the H$\beta$ line, together with an absorption
feature on the red side of the profile. Both features modulate 
with the stellar rotation and are nicely explained within the
framework of the X-celerator model of Shu et al. (1994 -- see also 
Johns \& Basri 1995a). 

The large majority of projects taking advantage of time series
observations use high resolution data. Thus, with unprecedented 
detail, information regarding the gas configuration of the regions
contributing to each individual feature of the line profile is
consistently recovered. The intrinsic limitations are to obtain a 
sufficient number of nights to secure the completion of the goals and 
to have
enough signal to recover the photospheric absorption lines which are
crucial for determining veiling.

We initiate a project based on time series
observations of CTTS, aiming to collect spectrophotometric and high resolution data of 
``VAriability of Southern T Tauri Stars'' (VASTT, henceforth). The main goal is to monitor the
spectral line variability and the continuum changes as a means of improving our understanding of their
surrounding circumstellar gas.  We attempt to recover
flux-calibrated observations, weather allowing.  When unfeasible, we
attempt to recover at least the shape of the spectral energy
distribution after observing photometric standards throughout the
night.  In this contribution we determine the evolution of
the H$\beta$ line emission component, its redshifted absorption
component and the spectral veiling for the star GQ Lupi, which is a
well known representative of the YY Ori sub--class.

\section{Observations}

The data presented in this communication are part of a larger data set
of CTTS observed during 6 and 8 consecutive nights of May and July 1998,
respectively.  The stars
were selected from the brightest members of the Cha and Lupus complexes
and a detailed study of each will be presented in forthcoming
papers.

The EW$_{H\alpha}$ of GQ Lupi is 2.8 \AA~ (Herbig \& Bell 1988) which
is certainly not typical for a CTTS. The ``Pico dos Dias Survey'' (PDS)
indicates a somewhat larger H$_{\alpha}$ equivalent width of 86 \AA~ (C.A.O.Torres -- private communication). The classification of
GQ Lupi as a member of the YY Ori sub--class together with its reported
variability led us to include this star in our survey of CTTS. However, 
the motivation for tackling the GQ Lupi data set first is based on the
appearance of inverse PCygni profiles seen conspicuously in the high
Balmer line members up to H$\beta$ and the CaII H and K lines  (see
Figure~\ref{fig1} in JD = 2450942.925). The IPC signatures vanish a 
few days
later (see Figure~\ref{fig1} in JD = 2450947.900) with the Balmer lines 
and the
Balmer discontinuity presenting their largest emission. Two months
later, our second observing run caught the strongest absorption of an
IPC ever reported at our resolution, with the H$\beta$ line indicating 
no more than a
modest peak of emission (see Figure~\ref{fig1} in JD = 2450997.752). 
July observations show smoothly varying line strengths.
Other CTTS showed inverse PCygni profiles during our campaign but
neither their variability nor their strength matches that of GQ Lupi.

GQ Lupi inverse PCygni profiles have been reported previously by
Bertout et al. (1982) who assign to this feature the highest variability
among all other line features. It is listed as a K7 V star, with
EW$_{H\alpha}$ = 2.8 (Herbig \& Bell 1988).

\subsection{Instrumentation and Data Reduction}
 
Spectroscopic observations were collected with the ESO 1.52m telescope 
at La
Silla with the Boller \& Chivens spectrograph. Spectral windows extend
from 5200 \AA~ down to 3100 \AA, which includes the major spectral
features directly responsible for the strong U and B modulation
reported for TTS.  Gratings are chosen to maximize the response 
(\# 31 and 33)
at these wavelengths. Having in mind that precise
veiling measurements are an important aspect of this program, we
aim for a S/N larger than 100 during clear nights.  
An integration time of 10 - 15
minutes for a star of $V=12$ was typically employed. The spectrograph
feeds a Loral Lesser CCD instrument (2048 x 2048)  providing an
effective resolution of 2 \AA. The slit was open to 5 arcsec to
include the whole seeing disk.  Some of the stars were monitored
several times a night (GQ Lupi included) and large zenith angles were
likely to occur. Thus, the slit orientation was moved to a new
position angle in order to correct for differential refraction and
minimize the light losses.  The corrections rely on the
curves presented by Filippenko (1982). Table 1 presents the observation
logs. Observers on preceding and proceeding nights joined
efforts to monitor the GQ Lupi spectral variability. The instrumental
settings were not necessarily the same but they allow a few overlapping 
spectral
windows, such as that including H$\beta$. This supplementary data
will be reported in another communication.

\figcaption[fig1.ps]{Spectrophotometric observations of GQ Lupi for the May and July (1998) runs. Only exposures representative of each observing night are presented. Conspicuous inverse PCygni profiles (IPC) are seen in some of the exposures, especially at JD = 2450997.752. Significant line emission is detected at JD = 2450947.900, with no indication of inverse PCygni absorption component.
\label{fig1}
}

The data is reduced in a standard way, with flat--fielding, bias and
sky subtraction, using IDL--based routines developed by G.
Basri (private communication). Due to the flexing  of the optical
system at different telescope positions, a calibration lamp (He--Ar)
is taken after each series of observations targeting a group of closely
spaced stars. We identify an average of 30 lines with a 5$^{th}$ order
polynomial fit, yielding an accuracy of about 15 km s$^{-1}$ in the dispersion solution.

\subsection {The Flux Calibrations}

Two spectrophotometric standards (Hamuy et al. 1992) were observed
throughout the night and used as flux calibrators (EG 274, LTT 7379).
Essentially, we observed rounds of 5 to 10 TTS and then the standards,
thus assuring a series of response curves for each night distributed
over different air masses. The nights were not photometric, prohibiting 
an accurate determination of the extinction law and subsequent flux 
calibrations.  Consequently, stellar radius cannot be derived in a 
self consistent way.
During the 1999 observing run, whose analysis will be presented in a
forthcoming communication, we observed GQ Lupi for nearly 20 
consecutive nights and under ideal conditions, which will permit 
us to access its basic stellar parameters.

We adopt the May and July (1997) average extinction
coefficients of La Silla, kindly provided by the Geneva Observatory
Photometric Group.  Variable atmospheric transmission on a daily or
yearly basis will change the absolute value of the adopted
coefficients -- which are wavelength dependent -- thereby affecting 
the final spectral energy distribution (SED) of the star. 
After inspecting the night to night
variability of the Geneva coefficients, we estimate these errors to be
less than 2 $\%$, which are negligible when compared to the intrinsic 
stellar variability.

Wavelength independent flux shifts are likely to be incorporated into
the nightly set of sensitivity functions (SF) due to cloud transit.
To monitor the shape changes of the TTS spectral distribution - 
which is one of our
goals - one needs to estimate the instrumental signature of the night.
The whole set of response curves are moved to the
brightest, which is assumed to be taken under the best sky conditions. 
Thus,
the residuals among the response curves are minimized after gray 
shifting each
curve to their new average.
The absolute flux error
due to cloud transit is the difference between the dimmest and the
brightest response curve and can be found in Table 1. The internal 
error of
the shifted SF's around the median is usually less than 4$\%$ for 
3400 \AA~
$< \lambda < 5000 $ \AA~ and rises steadily outside this region. At
wavelengths $< $ 3400  \AA,  the signal diminishes due to a drop in the
atmospheric transmission which leads to a poorly defined spectral light
distribution. We list in Table 1 the absolute flux uncertainties due to
variable atmospheric transparency. 

A subset of the observations, with representative spectra of each 
of the observing nights,
is displayed in Figure~\ref{fig1} with their corresponding
Julian dates.
Light losses were registered at  JD = 2450944.687 and 2450945.826,
especially in the blue, due to the transit of cirrus, severe winds,
 and the consequent
effect on the tracking. The corresponding SED's are therefore unreliable.

\section {Results}

Detailed identifications of the emission lines present in the GQ Lupi spectrum are presented in 
Appenzeller, Mundt \& Wolf (1977) and Bertout et al. (1982). It is possible to infer from the Bertout et al. 
(1982) data that this CTTS has moderate levels of continuum excess. The photospheric absorption lines are well 
discerned even in the blue part of the spectrum where the continuum excess is more severe. It is true that 
the veiling is better determined with high resolution observations. However, this particular spectral format
offers several deep lines which can yield average veiling values over 50 \AA~ windows (see below).

The Balmer emission lines -- markers of disk accretion (Valenti, Basri \& Johns 1993 -- VBJ henceforth) -- plus 
the H \& K CaII lines are the only consistent emission lines throughout the observing run. The helium line 
$\lambda$ 4471 and perhaps $\lambda$ 4026 are also marginally present. On the last night of the May run, 
the Balmer emission increases severely and the Fe II lines $\lambda\lambda$ 4923 and 5018 appear.

\subsection{The Time Series of the Spectral Continuum Distribution}

The variable transparency over the course of the observing run is a hindrance to 
directly measuring the SED. We therefore investigate the evolution of flux ratios 
instead, assuming achromatic flux shifts cancel out.  Spectral windows representative 
of the local continuum are defined for each of the observations  which 
are then normalized by the flux, F$_{4970}$.

\figcaption[fig2.ps]{Spectral windows clear of emission lines are selected for determining the Balmer and the Paschen continuum slopes. Each spectrum is normalized at 4970 \AA , and an offset equal to the time lag between consecutive exposures is
added for display. Light losses were registered for the observations plotted
with dot-dash lines.
\label{fig2}
} 

\subsubsection{The Balmer Jump}

The Balmer jump of several CTTS has been previously measured and compared with 
other properties of the gas flow (VBJ, Gullbring et al. 1988 -- GHBC, Calvet \& 
Gullbring 1998 -- CG). These works bring together single observations of a variety 
of CTTS while this study investigates
the evolution of the Balmer jump for a single star.  VBJ and CG compute the Balmer 
jump by modeling the excess emission flux (de-reddening the observed SED and subtracting 
out the spectral distribution of the underlying photosphere).  We take a slightly different
approach, quantifying the Balmer jump directly from the observed SED. More specifically, 
the Balmer jump is defined to be the ratio between the fluxes blueward and redward of the 
nominal discontinuity.  We assign 10 points that are far from major emission
lines or strong absorption lines to determine where the local continuum
is. The closest  {\it continuum} window redward of the Balmer jump is
at 3860 \AA~, 200 \AA~ off the nominal Balmer edge. The spectra are normalized
to the continuum flux window at 4970 \AA, and a constant equal to the time lag between
observations is added to these normalized spectra for presentation in 
Figure~\ref{fig2}. Light losses were registered during two of the exposures and 
the corresponding SED should be analyzed with caution (see the dot-dash data
of Figure~\ref{fig2}. 

The spectral distribution changes clearly on a night-to-night basis.  In particular,
variability is noted at frequencies near the Balmer discontinuity. 

Figure~\ref{fig3} illustrates the evolution of the flux ratio across the Balmer
jump ($\lambda$F$_{\lambda< 3640}$/$\lambda$F$_{\lambda>3640}$). Open circles 
indicate the data subjected to light losses.
Measurement of the Balmer jump is subject to various systematic errors.
The unresolved forest of blended emission lines (H and CaII) prevents a clear 
definition of where the continuum should rest.  The situation is also problematic 
when there is significant absorption in the inverse PCygni profiles, especially those 
near the Balmer discontinuity.  The nightly response curves are
uncertain for wavelengths shorter than 3400 \AA. Therefore, only 
200 \AA~ is effectively used to identify the Balmer continuum slope.
On the other side of the jump, 
200 \AA~ separates the last well-identified (Paschen) continuum
window from the edge of the jump, reducing the precision of a polynomial
extrapolation. The continuum window at 3860 \AA~ is
abandoned in the extrapolation procedure due to the presence of the CN band.
Thus, we make linear extrapolations of the closest continuum 
windows (two in the Balmer and four in the Paschen continuum) towards the jump. 
The extrapolated Balmer continuum flux deviates by less than 2 $\%$ of the 
observed flux near the edge, increasing to 5$\%$ in the spectra of lowest S/N.
Assuming a similar uncertainty in the projected Paschen continuum flux
towards the jump, the estimated error of the Balmer
jump flux ratio is less than 10$\%$ (two-sigma error bars are 
used in Figure~\ref{fig3}).

\figcaption[fig3.ps]{The Balmer jump time series is shown for the May and July runs. The Balmer jump is defined in section 3.1.1 as the flux ratio between the extrapolated Balmer and Paschen continua towards the jump. It shows smooth variability, with local maxima at JD = 2450947.911 and 2451003.473.
\label{fig3}
}

\figcaption[fig4.ps]{The GQ Lupi time series reveals large emission on the night beginning at JD = 2450947.607. The Paschen continuum excess does not change significantly while the Balmer continuum excess does (solid line). At JD = 2450997.511, the Balmer jump is  minimum (dots).
\label{fig4}
} 

\figcaption[fig5a.ps]{{\bf a)} The Balmer continuum slope time series as a function of the Balmer jump. Arrows indicate the time progression for the July series. 
{\bf b)} The same as {\bf a)}, for the Paschen continuum slope.
\label{fig5}. The data subjected to light loses are shown as crosses.
}

During the last night of the May run (May 14, 1998), the Balmer 
continuum brightens while the Paschen continuum
maintains its shape and strength. This can be seen in Figure 4 where
each spectrum is shown with solid lines.  Contrastingly, the Balmer
jumps of JD = 2450997.752 decrease to a minimum, the
Balmer slope being nearly a continuation of the Paschen continuum
towards shorter wavelengths.  This is illustrated in the same 
Figure as dots. We note the excess in the B
band seen in the curve of JD = 2450997.752 (broken line), which will be
further explained (see Figure~\ref{fig6} below) as a direct result of different
veiling distributions.  In this exposure, the Balmer emission lines
show the largest inverse PCygni profiles ever observed in GQ Lupi,
perhaps in any classical TTS (see Section 3.3.1).  Furthermore, sharp
U--band variations on the last night of May are also depicted in Figure~\ref{fig4}. We closed the May run with GQ Lupi showing a very large Balmer jump and
opened the July season with nearly no Balmer jump. It then evolves towards
values larger than 2.0, increasing and then fading on the last two
nights.

It is important to emphasize, in regards to Figure~\ref{fig4}, that the
brightening of the Balmer emission is not followed by a
concurrent change in the shape or strength of the Paschen
continuum (see the next section on the measured veiling). It suggests that
the brightening of the Balmer continuum comes from a region that is physically
more extended or not directly associated with that of the Paschen continuum. Density and/or temperature changes in the emitting region might 
also cause similar behavior.

The evolution of the Balmer jump depends  on the evolution of the Balmer and the Paschen continuum slopes. In Figure~\ref{fig5}  {\bf a},{\bf b} 
we correlate the Balmer jump with the
slopes of the Balmer and Paschen continua respectively. Open circles are used
for the May observations.  The reduced number of
observations in May precludes us from inferring any obvious evolutionary
trend in the slopes for this month. Arrows are included in both panels to indicate the
time progression of the exposures.  First, we note that the
Balmer jump is correlated with both the Balmer and Paschen 
continuum slopes (linear Pearson coefficients of -0.73 and 0.85 are computed
for the Balmer and Paschen slopes respectively). This is expected if the 
forming regions have a common energy source providing
the emission. Second, both slopes start out as if the temperature 
of the emitting region was progressively growing. If no excess emission is 
present, the slopes
should reflect those of a cool photosphere peaking at red wavelengths. The 
progression of the slopes indicate the brightening of the emission
region. The temperature increase shows cadence with the size of 
the  Balmer jump (ordinate).  However, the trend can also be explained
by combining different sizes of emitting areas, gas temperature and
opacity (the slab model scenario) or by advocating distinct emitting
regions for the Balmer and the Paschen continuum (the shock model scenario
(CG)).
This trend is reversed at JD = 2451003.473, when the Balmer jump starts
decreasing as the blue excess increases. Figure~\ref{fig5}
{\bf a} and {\bf b} also indicates that the Balmer continuum slope throughout the season changes less (0.2) than that of the Paschen continuum (0.5).

These changes are, in fact, artifacts of the spectral veiling distribution
(Hartigan et al.
1988; Basri \& Batalha 1990; Hartigan et al. 1991) which can be consistently
measured in GQ Lupi. 

\subsubsection{The Veiling Time Series}

Several absorption lines are present in this spectral format and are
well distributed over the region of the  Paschen and the Balmer
continua. They are used in previous calculations of VBJ and GHBC to
recover veiling in a large sample of TTS, observed with similar
spectral format. The CaI line at 4226 \AA~ is one of the
strongest absorption lines in the Paschen continuum region, followed by
a family of strong iron lines at $\lambda>$ 4800 and a TiO band at
4750 \AA,  not always strong in the GQ Lupi time series. The lines blueward of the Balmer
jump suffer from being in a region of lower S/N due to a combination of
decreasing instrumental response and less stellar flux. Nevertheless, a
strong blend (at our resolution) of several iron lines of low
excitation potential at 3550 \AA~ is clearly seen in most of the
data, allowing us to extend the {\it veiling distribution}, as a function
of time, to this region. The star Gliese 825 (K7 V) is the 
template for the de--veiling  procedure.

We divide the spectrum into windows of 50 \AA~ and assume that the
excess of continuum external to the photosphere (F{\it ext}) is flat 
within this window ($F_{ext}/F_{phot} = const$).
This is supported by previous measurements of
TTS veiling taken  with data of higher S/N (Stout--Batalha, Batalha \&
Basri 2000; Hartigan et al. 1991; Basri \& Batalha 1990). The spectra
of the TTS and the template are normalized over a larger spectral
region (150 \AA) in order to decrease the uncertainties regarding the
continuum placement. Then, a first guess to the average veiling is
obtained by minimizing the $\chi^2$ of the model fit (the veiled template) applied over the entire 150 \AA~
window. This is followed by fine adjustments (steps of 0.1 in veiling). We estimate veiling measures with errors less than $\pm$
0.15 at $\lambda > $ 4000 \AA.  The error bar is larger at 3550 \AA~ and depends on
the absolute veiling. This is partially due to poor continuum normalization in this region combined with the
effects of low instrumental response. It is usually $\pm$ 0.5, for values of
veiling less than 4.0, and increases to about  $\pm$ 1.5 for the largest
veiling  ($\sim$ 8.0). The blend at 3550 \AA~ is not detected on the last night of the May run,
preventing any veiling determination in this region.  

\begin{figure}
\caption{The veiling spectral distribution is shown for the subset of the GQ Lupi exposures of Figure 1 (dots). Several lines in the Paschen continuum assure a reliable veiling determination in this region. A blend of low excitation potential iron lines is the only probe for veiling in the Balmer continuum region.  During the night beginning at JD = 2450947.607, the excess
of Balmer continuum flux increases (see Figure~\ref{fig4}), preventing any veiling determination at $\lambda$ 3550 with our resolution. The model-veilings 
developed in section 5 are also indicated.}
\label{fig6}
\end{figure}

The complete veiling time series is shown in Figure~\ref{fig6}. Also shown 
are the model-veilings (full line) computed and discussed in section 4.0. 
The slope of the distribution and the absolute value of the excess at a given
wavelength change on short (hours) and long (days) time scales.  We
will explore the correlation of these changes with other
spectral features. For now, we emphasize the daily variations of the veiling
parameter, showing the change in absolute values at 4000
\AA~ and 5100 \AA~ (Figure~\ref{fig7} {\bf a},{\bf b}),  reflecting the variability of
the broad B band and the V band (our nearest data point). 
At the beginning of the May run (JD = 2450942.8846), we find GQ Lupi
in a relatively bright state, with an average veiling close to 2.5 in
the B band.  High cirrus prevented any science on the following night.
On JD = 2450944.6867 and the next night, the GQ Lupi veiling
is about half of this value. The shapes of the Balmer lines are similar
to those reported previously for YY Ori stars, with conspicuous inverse
PCygni profiles that are even visible in the red wings of the H \& K CaII lines
(see Figure~\ref{fig1}). The last night of the May run (beginning on JD =
2450947.6066), the veiling recuperates its original strength and the
Balmer lines experience the highest emission ever reported for this
star - H$\beta$ ranges between 20 and 30 \AA). We observe large variations of the continuum in less
than 6 hours.  

\begin{figure}
\caption{The veiling time series shows smooth variability. We indicate the flux excess at 4000 \AA~ {\bf a)} and 5000 \AA~ {\bf b)}, representing the
expected changes in regions near the B and V bands.}
\label{fig7}
\end{figure}

The July time series displays a steady and progressive
decrease with local maxima at JD = 2450997.8, 2451000.6 and 2451004.5.
The changes near the V band ($\lambda$ = 5100 \AA) are similar to those of
the B band but less accentuated, especially on the last night of the May
run.

The continuum excess at 3550 \AA~ relies upon a single measurement and
undergoes strong variability.  The July run shows a smooth increase in the Balmer jump, with a possible turn over by the end of the observing
run, a behavior closely anticorrelated with the veiling.  Figure~\ref{fig8} shows how tight this behavior is, especially for the July 
data set.  A similar relation was previously noted by VBJ and GHBC for a set of different
CTTS, though with larger scatter.   

\begin{figure}
\caption{The Balmer jump anticorrelation with the veiling is clearly depicted in this Figure.
The May observing run presents a scatter larger than that of the July run.
The Balmer jump vanishes for veilings larger than 4.5}
\label{fig8}
\end{figure}

\subsection{The Circumstellar Extinction}

The SED of a TTS is empirically modeled using an observed photospheric
template of the same spectral type and gravity, the
veiling distribution of the TTS, and the circumstellar extinction as
as free parameter. If the spectrophotometric data is taken during
clear nights, both the stellar radius and the circumstellar reddening
are consistently and elegantly derived (see GHBC).  Our observations were
taken under variable atmospheric transmission, which prevented us from 
determining the stellar radius -- a lower limit at best. Thus, we decide 
to compare only the shape of the empirically modeled SED to the shape of 
the observed SED.  The veiling spectral distribution directly determined 
from spectroscopy is
added to the template photosphere to compose the empirical systemic SED
of each exposure (see  Eq. [1] in
Hartigan et al. 1991). The circumstellar extinction is then inferred after
comparing an artificially reddened systemic SED with the corresponding
observation.
The standard interstellar extinction curve of Cardelli, Clayton \& Mathis (1989)
is adopted.

For a sufficiently large number of
exposures with well computed veiling distributions we thought that the
previous exercise would lead us to a single value of circumstellar extinction.
On the contrary, we were unable to converge to a single A$_v$ for one
or more of the following reasons.  First, the observed flux distribution 
does not match the modeled SED at $ \lambda <$ 4000 \AA.  VBJ also notes
mismatches between the continuum distribution of their slab models and the 
observations, within the wavelength range 3700 - 3900 \AA. They suspect that
errors introduced during the flux calibration procedure affect the final 
distribution of their adopted flux standard. Our largest source of error is 
due to a single veiling determination in the Balmer continuum (3550 \AA), 
that has the largest error bar in the format ($\pm$ 1.5). Thus,
extrapolations towards $\lambda < 4000 $ \AA~ are subject to large 
uncertainties. Second, and since our data is not flux calibrated, we force 
normalizations at a given wavelength in order to compare the shape of the 
modeled spectral energy distribution to the observed one. Different values 
of the circumstellar extinction parameter would prompt changes not only in 
the shape but also in the flux level of the semi-empirical model. Obviously,
our data prevent the latter test, reducing our ability to determine the
extinction. Third, the veiling distribution is subject to significant 
interpolation errors due to sparse sampling.
Finally, the small spectral range available to access A$_v$ 
($\Delta\lambda\sim 1500 \AA$) reduces the precision. 

Given the above uncertainties, we decide to compute the extinction using
the average SED of all of the de-veiled observations.  
We normalize the 32 de-veiled SED's to the brightest. The 
resulting curves share similar slopes and deviate less than 10$\%$ from 
the mean at $\lambda > $4000 \AA~, increasing up to 20$\%$, at 3300 \AA~. The 
mean curve is assumed to be the GQ Lupi reddened photosphere. The template
Gliese 825 (K7 V), is artificially reddened until a sufficient fit to the average 
reddened photosphere of GQ Lupi is attained. Our data covers a small spectral 
window, thereby limiting the precision at which the extinction can be inferred.
Nevertheless, this exercise leads us to an extinction value of A$_v$ = 0.4 
$\pm$ 0.2. In section 5.1, we develop models of the 
emitting region and establish an optimum model value within this range.

\subsection{The H$\beta$ Line Flux Time Series}

\subsubsection{The Absorption Component}

Two components of the H$\beta$ profiles are clearly distinguished: the
emission and the redshifted absorption component, which is indicative of
mass infall. The equivalent width  of the latter is a rough measure of
the column density of obscuring gas crossing the line of sight. Thus, 
we compute the
total depleted area in the extreme red wing of the H$\beta$ profile,
after subtracting the corresponding profile of an inactive photosphere.
The results are shown on the left hand side of  Figure~\ref{fig9}.  
We are
unable to verify any correlation between this absorption and the 
veiling or the
Balmer jump during the May run, perhaps due to the modest number of 
observations.
This perspective is clearly changed in July, however, when the maximum 
absorption
coincides with the maximum veiling in the beginning of the mission.
Both parameters decrease steadily up to the end of the
mission. Some kinks larger than the uncertainties in the veiling time 
series are visible at JD =
2451000.469 and 2451004.472 but are not reflected in the absorption 
component.
In the left panel of Figure~\ref{fig10}, we plot the strength of the 
continuum excess emission versus the H$\beta$ absorption component for
both the May and July data sets.
No trend can be established for the May data set. On the contrary, the July time series shows a clear correlation between the strength of the redshifted absorption and the veiling. Observations with large veiling (V$_\lambda >$ 3.0) 
show strong IPC. 

\begin{figure}
\caption{ a) The redshifted absorption component of the H$\beta$ line is measured and presented as a function of time (JD - 2450938). The May time series (left top panel) has a modest data set. On the last observing night, beginning at JD = 2450947.607, the redshifted absorption is not detected, and the emitting component is at a maximum. The absorption component shows regular behavior during the July time series (left bottom panel). The series starts with the absorption near maximum
and decreases steadily to a minimum at JD = 2451001.472, strengthening once more thereafter.
It should be emphasize that these EW are not corrected for veiling; if they were, the correlation would improve. b) The equivalent width of the H$\beta$ emission component, not corrected for the veiling continuum emission. The May time series (right top panel) shows the dramatic increase in line emission by the end of the run; the July time series, indicates the
steady strengthening of the emission component (right bottom panel).  }
\label{fig9}
\end{figure} 

\begin{figure}
\caption{a) The correlation between the veiling and the  absorption component is clearly established (left panel). It should be emphasized that if the absorption EW were properly de-veiled, the correlation would strengthen even more. b) On the contrary, the H$\beta$ emission component is anticorrelated with the excess of continuum emission (right panel, circles).  If the emission components are de-veiled, not an obvious change in line fluxes up to veiling of 1.0 is detected. For larger veiling, the line fluxes apparently diminish. }
\label{fig10}
\end{figure}
 
It should be emphasized that these EW's
are not corrected for the  continuum veiling. As commented
in Basri \& Batalha (1990) and Hartigan et al. (1991), the inclusion of such corrections would
increase the line profile by a factor proportional to the veiling
itself (1 + V$_\lambda$), which will further enhance the correlation. The spectrophotometry of Lehmann, Reipurth \& Brander (1995), done for EX Lupi,
also indicates that the maximum emission -- i.e. maximum veiling
-- coincides with a strong IPC.

The absorption component is completely washed  out during the five exposures
of May, beginning on JD = 2450947.607, while the emission components
show the strongest emission of the whole time series. Despite the absence of absorption, a column of obscuring material is still feeding the star, and
its presence is inferred by the asymmetric shape of the H$\beta$
emitting profile (see Figure~\ref{fig11}).  During this particular night, the
Balmer jump reaches the highest value of the 1998 -observing season
and increases along with the H$\beta$ strength. Indeed, theoretical predictions (Hartmann 1991) support the correlation between the optical veiling and the strength of the redshifted absorptions. Johns--Krull \& Basri (1997) attempt to establish this correlation in DF Tau high resolution time series with no conclusive results. 

\begin{figure}
\caption{The H$\beta$ lines show, in general, the redshifted absorption components typical of YY Ori stars. At JD = 2450947.607 (May 14 $^{th}$), the emission grows to a maximum
and no absorption is detected. Nevertheless, the red wings of the profile show clear signatures of depletion, indicating that mass transfer towards the star is still in operation.}
\label{fig11}
\end{figure} 

\subsubsection {The Emission Component}

To properly compute the line flux excess, which originates in
the circumstellar environment, the underlying photospheric profile which
is assumed to be constant, must be subtracted out.
Since our data are not
of photometric quality, we aim at flux ratios instead. We compute the
total emitting line flux
normalized to the local continuum. The complete range in variability impressed upon 
the H$\beta$ profile
during the 1998 observing season is displayed in 
Figure~\ref{fig11}.  It is
interesting to follow the progression of the absorption minimum
and the emission peak as the total emission grows.  Both cores
move redward as the H$\beta$ emission strengthens,
indicating the progressive depletion of obscuring infalling
material out of the line of sight. The strongest absorption coincides with 
the dimmest emission of the whole series. The
blue wing is less subject to variations.

The behavior of the H$\beta$ emission component as a function of time is
shown on the right hand side of Figure~\ref{fig9}. The line emission 
detected on the last night of May is unique among our observations.  
The variations 
seem to
be quite steady on the other nights. The anticorrelation between the 
emission and the absorption components (compare the left and right middle 
panel of 
Figure~\ref{fig9}) is striking - even the kinks at JD = 2451000.465 are 
seen in reverse.

The veiling is anticorrelated with the observed (not de-veiled)
line emission, here indicated by the nominal equivalent width 
(right panel of Figure~\ref{fig10}). However, to properly ascertain the dependency of 
the line emission on any of the parameters controlling the accretion properties 
of the gas, proper line de-veiling must be carried out.  The de-veiled emission 
equivalent widths are converted into line fluxes and displayed against the veiling  
in Figure~\ref{fig10}  (stars).  We do not find a clear dependence on the continuum 
excess flux, as found for the absorption component. There is a large scatter at 
intermediate veilings and the emission fluxes of May 14th are off scale. We note an 
apparent dimming of line emission at veilings larger than 3.20,  perhaps indicative 
of a real intrinsic weakening of line emission as veiling grows. However, the trend 
is not conclusive because of the low resolution of the data. The emission component 
is next to a variable and --for high veiling cases -- stronger  component that is in 
absorption. High resolution H$\beta$ time series should be taken in order
to confirm if the line emission properties -- which are primarily dependent on the 
circumstellar environment -- vary along with the changes in the shock region.

Changes in the veiling are governed by variable \.M or by changes 
in the field configuration which will affect the density and/or temperature of 
gas where the excess continuum emission originates. Figure~\ref{fig10} indicates that line emission 
is primarily ignoring whatever changes might be operational at the bottom of 
the 
accreting funnel. As veiling grows, so does the absorption along the 
line of sight that will translate into larger inverse PCygni absorption. 
The line flux emission, on the contrary, depends primarily on the 
hydrodynamic properties of the accreting gas, such as the local gas density and
temperature and the velocity field.

\subsection { The Brightening of GQ Lupi on JD=2450947.607 }

The smooth variability of the spectral features discussed herein is
met with one exception.  On JD = 2450947.607, most of the measured
parameters reveal their extremes: large veiling at 4000
\AA~, the largest Balmer jump, the largest Balmer and CaII emission
lines, along with the simultaneous appearance of the Fe II emission
lines.  The inverse PCygni profiles are not detected, and the largest
variability in the Balmer lines and Balmer continuum are also measured.
If veiling corrections are included to recuperate the real profile, the
H$\beta$  emission peak nearly doubles. This tendency is not verified on
other nights.

The strengthening of the Balmer continuum emission can be inferred by the
complete absence of the $\lambda$ 3550  blend of iron absorption lines. 
Thus, Balmer continuum veiling is not measured this night. 
The sudden flux increase of the Balmer continuum, as shown in 
Figure~\ref{fig4}, is not followed by a similar brightening of
the Paschen continuum.  The slope of the latter suffers
negligible changes (see the open circles Figure~\ref{fig5} {\bf a},{\bf b} ) while the jump
strengthens. In fact, the flux excesses displayed in Figure~\ref{fig6} reflect
this behavior: the average veiling distribution at $\lambda >$ 4000
\AA~ does not change with the sudden increase of the Balmer continuum, strongly 
supporting distinct formation regions for both spectral features on this 
particular night.  However, a common region of formation can be argued
if the Balmer jump increase is the result of a slightly higher gas
temperature and significantly higher filling factor (see Section 4.5).

\section{Discussion}

\subsection {The Balmer Jump $vs$ Veiling Anticorrelation}

The continuum excess flux anticorrelates with the Balmer
jump (Figure~\ref{fig8}). This trend was first reported by VBJ, using a large sample
of TTS and was interpreted within the slab scenario as a
competition between the H$^-$ and the H$_{b-f}$ opacities for a given
gas density and temperature. For low hydrogen densities, negligible
amounts of H$^-$ form, and the jump depends only on the
slab temperature.  As the density increases -- a direct result of
increasing mass infall or decreasing slab projected area -- the growing concentration of H$^-$ governs
the emission at the jump and eventually washes it out.

The anticorrelation is
also verified by GHBC, and they comment on the large optical
thickness of the emitting source to diminish
the jump. Nevertheless, they suggest that distinct regions might be
producing the
Balmer and the Paschen continua. Since stars of high accretion rate tend to
show negligible Balmer jumps, they conclude that the Paschen continuum
must increase faster than the Balmer continuum.

Shock models developed by CG
suggest that the Balmer excess is produced in the pre--shock region and is
optically thin. The flux excess along the Paschen continuum is optically
thick, arising in regions of the stellar atmosphere where the bulk of
accreting energy is reprocessed. The emission lines are formed preferentially
along the magnetic funnel if a certain configuration for the
magnetic field is assumed (Hartmann et al. 1994; Muzerolle,
Calvet \& Hartmann 1998). Allowing  different regions to contribute
to the final spectral distribution is an improvement over earlier slab
models -- especially if the asymmetric shapes and signatures of inverse
PCygni profiles are taken into consideration.  In this scenario, the veiling 
anticorrelation with the Balmer jump is understood if the Paschen continuum, 
which is governed by the mass infall, grows more rapidly than the Balmer 
continuum as the accretion increases.

\subsection{What Governs the Reported Variability?}

The accretion flow onto the central star affects several observables in
the optical range: the excess continuum of the Balmer and Paschen
regions, the total emission line fluxes, the shape of the line profile
and the strength of the absorption feature of the (inverse) PCygni
profiles. One of the primary motives in time series observations of pre--Main
Sequence stars is to determine the extent to which the emitting
quantities present periodic behavior comparable to the photometric periods of the central star, 
when available. Thus, we turn our attention to previous photometry of GQ Lupi. This star does not 
yet have a photometric period set in the literature.
Covino et al. (1992) publish light curves taken over 13 nights, but the
time sampling is not adequate for a firm period determination. We note,
however, that the local V band flux minimum repeats its approximate
value about 12 days latter.  P.C.Pereira (private communication) also
notes similar behavior after analyzing his unpublished photometry.  One can understand the 
difficulties involved in
establishing periodic behavior of GQ Lupi by observing the
veiling time series depicted in Figure~\ref{fig7} {\bf a},{\bf b}: an increase of about 1 mag
in the B band during the 8 nights of July, with two local maxima (in
flux). We submit the veiling time series to a periodogram analysis with no
conclusive results. If the stellar magnetosphere is
bipolar, the accreting gas will fall at high latitudes. Since GQ Lupi
shows inverse PCygni profiles in general, it is feasible to assume that
the system is observed with the line of sight nearly aligned with the
infalling column of gas, and that a large fraction of the reprocessed
energy of the accretion shock remains in view as the star rotates. The
major determinant in the changes are the field line configurations and/or 
the variable mass accretion resulting from the disk inhomogeneities. We develop 
models of the emitting region to identify which of these possibilities governs the continuum and 
the line changes.

\subsection {Slab Models and the Continuum Emission}

We approach the continuum emitting region with the simplest scenario to
treat a smooth flux continuum change: a slab of uniform density and
temperature. The observed variability is governed by changes in
the temperature, density or geometrical scale of the slabs.  Previous
slab models have explained some of the observations of a large set of
CTTS (see VBJ). GHBC computes slabs with particle densities of 
N$_H$ = 10$^{14}$
cm$^{-3}$ and three different regimes of opacity ($\tau_{3640} \sim $
0.1, 1.0, 10.0, respectively). They show that optically thick slabs of
high temperatures ($\sim$ 10000 K) are compatible with the observed
Balmer jump and the Paschen continuum slope, but inconsistent with the
observable Balmer continuum slope. The latter indicates gas emitting
regions of smaller temperatures. A natural improvement over slab models
is the accretion shock model of CG in which
the Balmer and the Paschen slopes as well as the Balmer jump are simultaneously
matched.  Contrary to the basic premise of slab models, their accretion shock 
models establish different regions for each of the continua. While the Paschen
continuum emission is kinetic energy released and reprocessed in the
after shock region, the Balmer continuum originates in the optically
thin gas of the pre-shock region in the upper photosphere. 
Our choice for using a simple approach is to determine which of the key
gas parameters governs the flux changes of the time series and not to 
develop precise models.

The general code CLOUDY is adopted to compute the radiation field
of such a slab.  The choice of CLOUDY is based on its simplicity and
flexibility in the choice of the external velocity field,
unimportant for calculations of the continuum flux but
important in the calculations of line formation. NLTE populations are
computed and input into an independent code to compute the line
profiles (Lopes \& Batalha -- in preparation). The standard continuum
opacity sources are included, including that of H$^-$. For each
hydrogen density, the statistical equilibrium equations of a 15--level 
atom are solved, and all collisional and radiative processes are included.  The
composed systemic SED is F$_\lambda$ = ( 1- $\delta$ )F$_{ph}$ + $\delta$ ( F$_{ph}$e$^{-\tau}$ + F$_{slab}$ ), where $\delta$ is the projected slab
area (filling factor in $\%$ of the stellar full disk area), F$_{ph}$ and F$_{slab}$ are the photospheric and the slab flux
distributions respectively, and $\tau$ is the slab optical depth.

We adopted the newly released version in C,  that provides reliable results 
for the range of densities and temperatures explored in this paper. 
We extend the gas densities up to blackbody limits for a characteristic
temperature and allow the geometric
extent of the slab to reach 1500 km (steps of 500 km), 
about twice the average thickness
of all models of VBJ. A grid of slabs models with densities larger than
N$_{H}$ $> 10^{13} cm^{-3}$, temperatures within the range 
5000 K $< Te < $12000 K, in steps of 100K, are computed, and the resulting
distribution, F$_\lambda$, compared directly to the de-reddened data. 
We adopt filling factors starting
at 0.05$\%$ and extending up to 20 $\%$ of the stellar surface.  For
densities less than 10$^{14}$ cm$^{-3}$, the Balmer jump of the slab,
diluted by the photospheric radiation, is larger than any of those
observed. The impact of the H$^-$ ion in the total opacity budget is
negligible and the opacity is controlled by hydrogen bound--free
transitions. As the gas density increases, the growing collisional
rates  eventually govern the statistic equilibrium, causing the
spectral energy distribution to approach a Planck function. The Balmer
continuum meets the blackbody distribution first, followed by the
Paschen continuum.  

\subsection {Modeled Slab Parameters}

Allowing all possible free parameters to float (temperature, slab dimensions 
and gas density), we search for a combination of gas parameters  providing
optimal fits to the observed SED's (the Balmer jump and the slopes
of the Balmer and the Paschen continua).  The circumstellar extinction is
constrained to A$_v = 0.4$ mag (see section  3.2). 
Each SED is defined by 14 continuum windows, 
including one at 3860 \AA~ that constrains the apparent absorption blend of the 
photospheric CN,  CaII H \& 
K and hydrogen outer wings. Although we are able to
obtain sufficiently good fits to the observed SED's, the models do not 
reproduce the observed veiling distribution for the exposures of
large Balmer jumps.  More specifically, the models 
underestimate the veiling at 4000 \AA, even if extinctions larger than 0.4 
mag are considered. By increasing the circumstellar extinction, hotter slabs have
to be considered to compensate for the depleted blue continuum, thereby
pumping the blue veiling. However, models of slightly larger chi-squared,
discarded during the model search procedure, are found to
provide simultaneous fits to the SED and to the veiling distribution.  
Final solutions are therefore chosen to be those which fit both the
SED and veiling distribution.  The final slab parameters are listed in 
Table 2 along with the relative standard deviations about the final fits.
The time series of the model parameters are
displayed in Figure~\ref{fig12}, and the actual fits to the sub-set of 
observations displayed in Figure~\ref{fig1} are shown in Figure~\ref{fig13}.  
Final model veilings are plotted in Figure~\ref{fig6}.  As mentioned in 
Section 2.2, light losses were registered on
JD = 2450944.687 and JD = 2450945.826, thereby affecting the observed SED.
Those model parameters are displayed with open circles in Figure~\ref{12}
and Figure~\ref{14}.

We emphasize the importance of using the veiling distribution 
to constrain models of continuum emitting regions as degenerate solutions
appear to be problematic (see VBJ).  
Knowledge of the veiling in a small wavelength 
window (e.g. at $\lambda 4226$ \AA -- a commonly 
used region for veiling determination) may result in degenerate solutions
as well.

\begin{figure}

\caption{ The slab model time series indicate a smooth variation of
parameters on a night-to-night basis. Open circles indicate nights
in which light loses were registered}
\label{fig12}
\end{figure}

\begin{figure}
\caption{Slab models are computed for the sub--sample of the GQ Lupi exposures presented in Figure~\ref{fig1} (dotted line). The best model (filled line)  provides the minimum  residual to the observation and fits the observed veiling. The geometrical thickness of the slab is found to be 1000 km, and the best model is chosen
out of a grid of varying N$_{H}$ (1.0 10$^{13}$ -- 1.0 10$^{17}$), Te (5000 K -- 16000 K) and filling factors (0.0005 -- 0.20). The model parameters are listed in each panel. The corresponding model--veiling are displayed in 
Figure~\ref{fig6}.}
\label{fig13}
\end{figure}

\subsection {Correlations Among Modeled Parameters}

We analyze the correlations among the model parameters to infer the dynamics 
of 
the region where the excess continuum originates.  The slab temperature 
governs 
the slope of the Balmer continuum while the slab density (and to a lesser
degree the slab projected area ($\delta$)) constrains the size of the Balmer 
jump.
Temperature and density are anticorrelated (Figure~\ref{fig14}a), and the total
emitting area increases with the gas density.  That is, models yielding a low
density slab of higher temperature have small emitting areas.  Increasing the 
density of a slab of a given temperature enhances the excess emission, and the 
distribution approaches the blackbody. Since the observed veiling distribution 
and the Balmer jump both constrain the models, one has to decrease the slab 
temperature to keep the model flux excess within the observed values 
if the gas density grows. 

The filling factor is the total projected area where the kinetic energy of the 
accretion is released. It grows for larger densities and lower temperatures. 
We interpret this growth as a direct consequence of mass accretion towards the 
central star, where the energy released in the after shock region is imparted 
over larger areas. 

The trends between the veiling and the model parameters are shown in 
Figure~\ref{fig14} {\bf b}. Large optical continuum excesses are the 
result of large gas densities spreading over larger areas.  Low veiling 
observations yield slab models of small projected emitting areas and larger 
temperatures.  The anticorrelation between the veiling and the Balmer jump 
(Figure~\ref{fig8}) is driven by density changes: the $\tau_{Balmer}$ 
is already
opaque, and the Balmer continuum distribution follows that of a blackbody of 
a given temperature (slope of Balmer continuum determines slab temperature). 
As  $\tau_{Paschen}$ increases, the Paschen spectral distribution merges with 
the blackbody distribution, which is an extension of the Balmer continuum.  
Veiling increases, and the
jump gradually disappears. The models indicate that the largest veiling states
of GQ Lupi have filling factors larger than the series average and slab
temperatures lower than the series average. 

Based on the nearly constant presence of IPC in the Balmer lines, 
it is possible to apply the magnetic field accretion scenario  
(Calvet and Hartmann 1992) to mediate the gas accretion and 
to establish where the continuum emission originates. 
The model parameters should indicate where the slab rests. Assuming that the 
gas is free--falling from a certain position R$_i$ in the circumstellar disk, 
the resulting disk accretion rate is (CG. see their eq. [6]):

$$
(\frac{\dot{M}}{10^{-8} M\odot yr^{-1}}) = (\frac{N_H}{5.2 10^{12}}) 
(\frac{M}{0.5M\odot})^{1/2} (\frac{R}{2R\odot})^{3/2} (\frac{\delta}{0.01}) 
(1 - \frac{R\star}{R_i})^{1/2}
$$

The computed number density for the highest veiling exposure at JD = 2450997.7525
is N$_{H}$ = 2.2 10$^{15}$ cm$^{-3}$. The resulting accretion rate is
larger than 10 $^{-5} M\odot$ yr$^{-1}$ which is too high for a classical TTS 
of modest veiling such as GQ Lupi. This density is that of a stellar 
photosphere 
with a mass column of 0.2 gr cm$^{-2}$, characterized by temperatures in 
the range 6000 - 9000 K.  Thus, the slab is not located in the accreting gas 
funnel between the circumstellar disk and the shock region.  Rather, the mass 
column is located in the photosphere adjoined to the base of the magnetic 
funnel, where the accretion energy is released.

The accretion luminosity (L$_{acc}$) is proportional to $\delta$F, where
F is obtained by integrating F$_{ext}$ (see Figure~\ref{fig6}) over
the entire spectrum.  Figure~\ref{fig14} indicates that the veiling 
(F$_{ext}$/F$_{phot}$) grows in cadence with the filling factor ($\delta$), 
suggesting that the accretion luminosity increases with the veiling.
This leads us to conclude that the progressive decrease of
the accretion rate ($\dot{M}$ $\propto$  L$_{acc}$) 
during the July time series governs the 
changes in the veiling and IPC absorption. The weakening $\dot{M}$,
however, does not prevent the reprocessed atmospheric gas from heating,
perhaps indicating that more of the region near the footpoint 
of the accretion column is being progressively exposed.

We look for other trends between the temporal behavior of the slab parameters 
and spectral features that are clearly produced in a physically detached region.
The redshifted absorption component, for instance, probes the infalling gas 
outside the atmosphere while the optical veiling is the reprocessed energy of 
the gas accretion onto the atmosphere. The redward absorption, which is 
dependent on the number of absorbers along the line of sight, is large for 
large values of $\delta$ (or $\dot{M}$)  and  weakens for decreasing 
$\delta$ (see Figure~\ref{fig9} and Figure~\ref{fig12} in the July run). 
Due to the YY Ori aspect
of the GQ Lupi line profiles, one can speculate that GQ Lupi is observed 
with the line of sight nearly aligned with the funnel axis. Thus, for larger
continuum emitting areas ($\delta$), more of the infalling gas will absorb
and deplete the underlying continuum. This trend, however, is not seen on the 
last night of the May run. The total emitting area is among the largest we 
model
and we detect no redshifted absorption. It is possible that the strong 
hydrogen 
emission on this particular night overcomes the absorption and fills it in. 
Theoretical efforts of Hartmann et al. (1994) predict that inverse
PCygni profiles will occur if an emitting region hotter than the photosphere
lies at the base of the accreting column of gas. Surely, the absorption we 
measure depends on the shape and emitting characteristics of the slab.

Given the fact that the Paschen continuum near the jump is not completely thick,
emission line components will necessarily be formed in the slab. Such emission 
lines were formerly modeled in VBJ. However, and especially for the low Balmer 
lines (H$\beta$ in our case), the core of the emission will be preferentially 
produced external to the slab. For H$\beta$, we note that the emission component
varies less than the redshifted absorption or the veiling (see 
Figure~\ref{fig10}), indicating that the bulk of the emission is governed by the
morphology of the magnetic field lines and the state of the controlled gas.

\begin{figure}
\caption{{\bf a} The temperature and the
density distribute themselves along a perfect locus. The filling factor
grows towards low temperatures.{\bf b} The correlation between the veiling and the model parameters.}
\label{fig14}
\end{figure}

\section{Summary}

We have presented time series spectrophotometric observations of GQ
Lupi, a well known member of the YY Ori sub-class of T Tauri stars
characterized by inverse PCygni profiles in the Balmer lines -- a clear
signature of mass infall. The sample comprises a total of 32 exposures
distributed over 13 nights in May and July 1998. In spite of the low
resolution of the data, we recover the excess flux distribution (i.e. veiling)
throughout the region of the Paschen continuum and, in most of the
cases, in the Balmer continuum at 3550 \AA~.

We summarize the major observational results as follows

1) The strength of the Balmer jump, defined as the flux ratio between
the extrapolated Paschen and Balmer continua towards the jump,
presents, in general, smooth variability (see Figure~\ref{fig3}).

2) The veiling time series also shows smooth and steady variability on a 
night-to-night basis (see Figure~\ref{fig6} {\bf a}, {\bf b}), and the largest
flux excesses are measured in the Balmer continuum. We confirm the previous findings of VBJ and GHBC, based on a large sample of CTTS, that the veiling anticorrelates with the Balmer jump (see Figure~\ref{fig8}). 

3) The constant presence of inverse PCygni profiles is the major characteristic of a YY Ori star.  
This fingerprint disappears from all of the Balmer and the CaII H \& K lines on the last night of 
the May run (1 out of 13 nights). Nevertheless, signatures of mass infall towards the central star 
are impressed in the emission lines (see Figure~\ref{fig11}).

4) The observed and not de-veiled equivalent widths of the H$\beta$ redward absorption are correlated 
with the excess continuum emission (see Figure~\ref{fig10}, left). 
On the contrary, the de-veiled H$\beta$ emission flux components do not change significantly for 
veiling less than 1.1 (V${\lambda}$ = 4000 \AA~), although large scatter is verified at intermediate 
veiling. The emission weakens
at larger veiling (see Figure~\ref{fig10}, right, and crosses).

5)  The apparently modest variability of the GQ Lupi continuum is not
a rule. In a series of observations taken on JD 2450947.607, the Balmer 
continuum shows flux excesses which grow to larger than 10
times the photospheric value, and the Balmer lines brighten well
above the average value of the series. The Balmer jump is at a maximum 
this particular night.

We model the observed continuum emission assuming it incorporates contributions 
from the stellar photosphere [(1 - $\delta$) F$_{\it ph}$] and from a slab of 
uniform temperature, particle density and height [$\delta$(F$_{\it slab} + 
F_{\it ph} e^{-\tau}$)], where $\delta$ is the projected area in $\%$ of the 
stellar photosphere and $\tau$ is the slab opacity. The 
circumstellar extinction for GQ Lupi is found to be A$_{v}$ = 0.4. 
The computed 
models reproduce the Balmer and the Paschen continuum slopes, the Balmer jump 
(see Figure~\ref{fig12}) and the veiling distribution (see Figure~\ref{fig6}). 
We summarize the major results as follows:

1)  The slab temperature governs the slope of the Balmer continuum, while the 
gas density and filling factors determine the size of the jump and the slope 
of the Paschen continuum. Thus, changes in the mass column density of the 
slab -- not in the gas temperature -- govern the veiling and the size of the 
Balmer jump.  

2) The slab temperatures are strongly anticorrelated with the slab gas 
densities, defining a perfect locus. The slab projected area increases with the density. We position
the slab in the stellar atmosphere near the shock region, where the free falling
gas reaches the stellar surface.

3) Models of large densities, large emitting areas and low temperatures fit all 
the data characterized by H$\beta$ with strong inverse PCygni profiles. In 
general, models of lower density and larger temperature fit data of mildly 
redshifted absorption and large Balmer jump.  

The slabs are optically thin redward of the Balmer jump.  Consequently, 
some component of the higher Balmer lines are likely to be produced in 
the slab.  For the lower Balmer lines, most of the emission originates in 
either a magnetically mediated accretion flow or in an expanding gas. The 
H$\beta$ lines show inverse PCygni profiles indicating their circumstellar 
origin.  When the observed H$\beta$ redward absorption is largest, the 
modeled slab density is at its highest for the series. The density steadily 
decreases as the absorption component weakens. The optical veiling is
tightly correlated with the H$\beta$ redward absorption strength.
The emission component is much more steady throughout the series, showing 
less dependence on the physical conditions of the post-shock region.

The atmosphere near to the shock region responds to the diminishing accretion 
rate ($\dot{M}$), showing less veiling and being confined to
smaller areas ($\delta$). On the contrary, the gas temperature grows as
does the observed Balmer jump. Additional spectrophotometric and high
resolution data of GQ Lupi are currently being analyzed to investigate 
how these trends evolve with time.

\begin{acknowledgements}

We thank G. Ferland for making available the latest version of CLOUDY before
releasing it.  We thank M. Fernandez for the careful reading of an earlier 
version of this paper. N.B. acknowledges the support of the Funda\c c\~ao 
de Amparo \`a Pesquisa do Estado do Rio de Janeiro (FAPERJ). 

\end{acknowledgements}

\begin{deluxetable}{ccccc}
\tablewidth{0pt}
\tablecaption{1998 Observing Log of GQ Lupi \label{obs}}
\tablehead{
\colhead{JD (mid.exp.)} &
\colhead{date}&
\colhead{weather transp. \%} &
}
\tablecolumns{4}
\startdata
2450942.8846 &5/09 & 20     & \nl
2450942.9252 &5/09 &            & \nl
2450944.6867 &5/11\tablenotemark{@} & $>$40 & \nl
2450945.8255 &5/12\tablenotemark{@} &            & \nl
2450947.6066 &5/14 & 35     & \nl
2450947.6748 &5/14 &            & \nl
2450947.8357 &5/14 &            & \nl
2450947.9003 &5/14 &            & \nl
2450947.9108 &5/14 &            & \nl
2450997.5107 &7/02 & 20     & \nl
2450997.6179 &7/02 &		& \nl
2450997.7525 &7/02 &		& \nl
2450998.4735 &7/03 & 17	& \nl
2450998.6319 &7/03 & 		& \nl
2450998.7088 &7/03 &		& \nl
2450998.7810 &7/03 &		& \nl
2450999.4673 &7/04 & 29	& \nl
2450999.6339 &7/04 &		& \nl
2450999.7539 &7/04 &		& \nl
2451000.4648 &7/05 & 15	& \nl
2451000.6232 &7/05 &		& \nl
2451001.4743 &7/06 &$>$40 & \nl
2451001.7369 &7/06 &		& \nl
2451001.7786 &7/06 &		& \nl
2451002.6388 &7/07 & 40	& \nl
2451002.6922 &7/07 &		& \nl
2451002.7618 &7/07 &		& \nl
2451003.4727 &7/08 & 35     & \nl
2451003.6624 &7/08 &		& \nl 
2451003.7468 &7/08 &		& \nl
2451004.4716 &7/09 & 27	& \nl
2451004.7120 &7/09 &		& \nl
\enddata
\tablenotetext{@}{Light losses were registered (see text)
and the SED is not reliable}
\end{deluxetable}

\begin{deluxetable}{rrrrr}
\tablewidth{0pc}
\tablecaption{Slab Model Parameters}
\tablehead{
\colhead{JD (mid.exp.)} &
\colhead{N$_{H}$ \tablenotemark{*} } &
\colhead{T (K)} &
\colhead{$\delta$ ($\%$)} &
\colhead{ $\chi ^2$ \tablenotemark{**} } 
}
\tablecolumns{5}
\startdata
2450942.8846 & 3.50 & 7200 & 5.2 & 0.92 \\
2450942.9252 & 1.10 & 8000 & 5.6 & 1.86 \\
2450944.6867\tablenotemark{@} & 9.90 & 6100 & 8.6 & 0.87 \\
2450945.8255\tablenotemark{@} & 1.60 & 7800 & 2.8 & 0.88 \\
2450947.6066 & 4.10 & 6700 & 9.4 & 1.07 \\
2450947.6748 & 5.50 & 6700 & 9.9 & 2.86 \\
2450947.8357 & 3.20 & 6900 & 9.0 & 1.37 \\
2450947.9003 & 1.20 & 7800 & 5.4 & 2.48 \\
2450947.9108 & 0.68 & 8200 & 3.6 & 1.36 \\
2450997.5107 & 1.40 & 7900 & 5.0 & 1.22 \\
2450997.6179 & 2.20 & 7600 & 5.6 & 0.62 \\
2450997.7525 & 2.20 & 7600 & 5.6 & 0.62 \\
2450998.4735 & 1.90 & 7700 & 5.0 & 1.59 \\
2450998.6319 & 2.10 & 7600 & 5.0 & 2.00 \\
2450998.7088 & 1.10 & 8100 & 4.2 & 0.79 \\
2450998.7810 & 1.00 & 8200 & 4.4 & 0.11 \\
2450999.4673 & 1.40 & 7900 & 3.0 & 0.32 \\
2450999.6339 & 1.80 & 7700 & 2.4 & 0.68 \\
2450999.7539 & 1.60 & 7800 & 2.4 & 0.88 \\
2451000.4648 & 1.20 & 8000 & 3.8 & 0.66 \\
2451000.6232 & 0.54 & 8600 & 3.8 & 0.47 \\
2451001.4743 & 0.38 & 9000 & 2.2 & 1.85 \\
2451001.7369 & 0.35 & 9100 & 1.4 & 1.96 \\
2451001.7786 & 0.34 & 9100 & 1.6 & 3.04 \\
2451002.6388 & 0.51 & 8600 & 1.8 & 0.72 \\
2451002.6922 & 0.47 & 8700 & 1.6 & 1.57 \\
2451002.7618 & 0.41 & 8800 & 1.6 & 1.88 \\
2451003.4727 & 0.37 & 8900 & 1.0 & 3.50 \\
2451003.6624 & 0.40 & 8700 & 1.2 & 2.03 \\ 
2451003.7468 & 0.42 & 8800 & 1.6 & 2.39 \\
2451004.4716 & 2.00 & 7500 & 3.0 & 0.55 \\
2451004.7120 & 1.20 & 7900 & 2.4 & 0.39 \\
\enddata
\tablenotetext{*}{N$_H$ in unities of 10.$^{15}$ cm $^{-3}$.}
\tablenotetext{**}{ $\chi^2$ is defined as 10.$^4$ (F$_{mod}$ - F$_{obs}$)$^2$/F$_{obs} ^2$. F is Log(flux). Acceptable solutions have $\chi^2$ less
than 50.}
\tablenotetext{@}{Light losses were registered (see text)
and the SED is not reliable.}

\end{deluxetable}


\begin{thebibliography} {}

\bibitem[]{} Aiad, A., Appenzeller, I., Bertout, C., Isobe, S., Shimizu, M., Stahl, O., Walker, M. F., \& Wolf, B. 1984, A\&A, 130, 67

\bibitem[]{}  Appenzeller, I., Chavarria, C., Krautter, J., Mundt, R. \& Wolf B. 1980, A\&A, 90, 184

\bibitem[]{} Appenzeller, I., Mundt, R., \& Wolf, B. 1977, A\&A, 63, 289
 
\bibitem[]{}  Appenzeller, I., \& Dearborn, D. S. R. 1984, ApJ, 278, 689

\bibitem[]{}  Appenzeller, I., Jankovics, I., \& Krautter, J. 1983, A\&AS, 53, 291
 
\bibitem[]{}  Appenzeller, I., \& Wolf, B. 1977, A\&A, 54, 713

\bibitem[]{}  Basri, G., \& Bertout, C. 1989, ApJ, 341, 340 

\bibitem[]{}  Bertout, C., Basri, G., \& Bouvier, J. 1988, ApJ, 330, 350

\bibitem[]{}  Bertout, C., Carrasco, L., Mundt, R., \& Wolf, B. 1982, A\&AS, 47, 419

\bibitem[]{}  Calvet, N., \& Gullbring, E. 1998, ApJ, 509, 802 (CG)

\bibitem[]{}  Calvet, N., \& Hartmann, L. 1992, ApJ, 386, 239

\bibitem[]{}  Camenzind, M. 1990, Reviews in Modern Astronomy (Springer-Verlag: Berlin), 3, p. 234

\bibitem[]{} Cardelli,J.A.,  Clayton,G.C., \& Mathis,J.S. 1989, ApJ, 345, 245

\bibitem[]{}  Covino, E., Terranegra, L., Franchini, M., Chavarria, K., \& Stalio, R. 1992, A\& AS, 94, 273

\bibitem[]{}  Edwards, S. 1979, PASP, 91, 329

\bibitem[]{}  Filippenko, A. V. 1982, PASP, 94, 715

\bibitem[]{}  Giampapa, M. S., Basri, G., Johns, C. M., \& Imhoff, C. L. 1993, ApJS, 89, 321

\bibitem[]{}  Grasdalen, G. 1977, Proceedings of IAU Colloquium 42, eds. R. Kippenhahn, J. Rahe, and W. Strohmeier, p.25

\bibitem[]{}  Gullbring, R., Hartmann, L., Briceno, C., \& Calvet, N. 1998, ApJ, 492, 323 (GHBC)

\bibitem[]{}  Gullbring, E., Petrov, P. P., Ilyin, I., Tuominen, I., Hackman, T., \& Loden, K. 1996, A\&A, 314, 835

\bibitem[]{}  Jacoby, G. H., Hunter, D. A., \& Christian, C. A. 1984, ApJS, 56, 257

\bibitem[]{}  Johns, C. M., \& Basri, G. 1995a, ApJ, 449, 341

\bibitem[]{}  Johns, C. M., \& Basri, G. 1995b, AJ, 109, 2800

\bibitem[]{}  Johns--Krull, C. M., \& Basri, G. 1997,  ApJ, 474, 433

\bibitem[]{}  Hamuy, M., Walker, A .R., Suntzeff, N. B., Gigoux, P., Heathcote, S. R., \& Phillips, M. M. 1992, PASP, 104, 533

\bibitem[]{}  Hartigan, P., Kenyon, S. J., Hartmann, L., Strom, S. E., Edwards, S., Welty, A. D., \& Stauffer, J. 1991, ApJ, 382, 617

\bibitem[]{}  Hartmann, L., Calvet, N., Avrett, E. H., \& Loeser, R. 1990, ApJ, 349, 168

\bibitem[]{}  Hartmann, L., \& Calvet, N. 1992, ApJ, 386, 239

\bibitem[]{}  Hartmann, L., Hewett, R., \& Calvet, N 1994, ApJ, 426, 669

\bibitem[]{}  Herbig, G. H., \& Bell, K. R. 1988, Lick Obs. Bull. 1111

\bibitem[]{}  K\"{o}nigl, A. 1991, ApJ, 370, L39

\bibitem[]{}  Krauter, J., \& Bastian, U. 1980, A\&A, 88, L6

\bibitem[]{}  Lago, M. T. V. T., \& Gameiro, J. F. 1998, MNRAS, 294, 272 

\bibitem[]{}  Lehmann, T., Reipurth, B., \& Brandner, W. 1995, A\&A, L9

\bibitem[]{}  Lopes, D. F., \& Batalha, C. 1999, in preparation

\bibitem[]{}  Mundt, R. 1983, ApJ, 1984, 280, 749

\bibitem[]{}  Mundt, R. 1984, ApJ, 280, 749

\bibitem[]{}  Muzerolle, J., Calvet, N., \& Hartmann, L. 1998, ApJ, 492, 743

\bibitem[]{}  Petrov, P. P., Gullbring, E., Ilyin, I., Gahm, G. F., Tuominen, I., Hackman, T., \& Loden, K., 1996, A\&A, 314, 821

\bibitem[]{}  Shu, F., Najita, J., Ostriker, E., Wilkin, F., Ruden, S., \& Lizano, S. 1994, ApJ, 429, 781 

\bibitem[]{}  Stout -- Batalha, N.M., Batalha, C., \& Basri, G. 2000, in press.

\bibitem[]{}  Uchida, Y., \& Shibata, K. 1984, PASJ, 36, 105

\bibitem[]{}  Valenti, J. A., Basri, G., \& Johns, C. M. 1993, AJ, 106, 2024 (VBJ)

\bibitem[]{}  Walker, M. F. 1972, ApJ, 175, 89

\bibitem[]{}  Wolf, B., Appenzeller, I.,  \& Bertout, C. 1977, A\&A, 58, 163

\end{thebibliography}
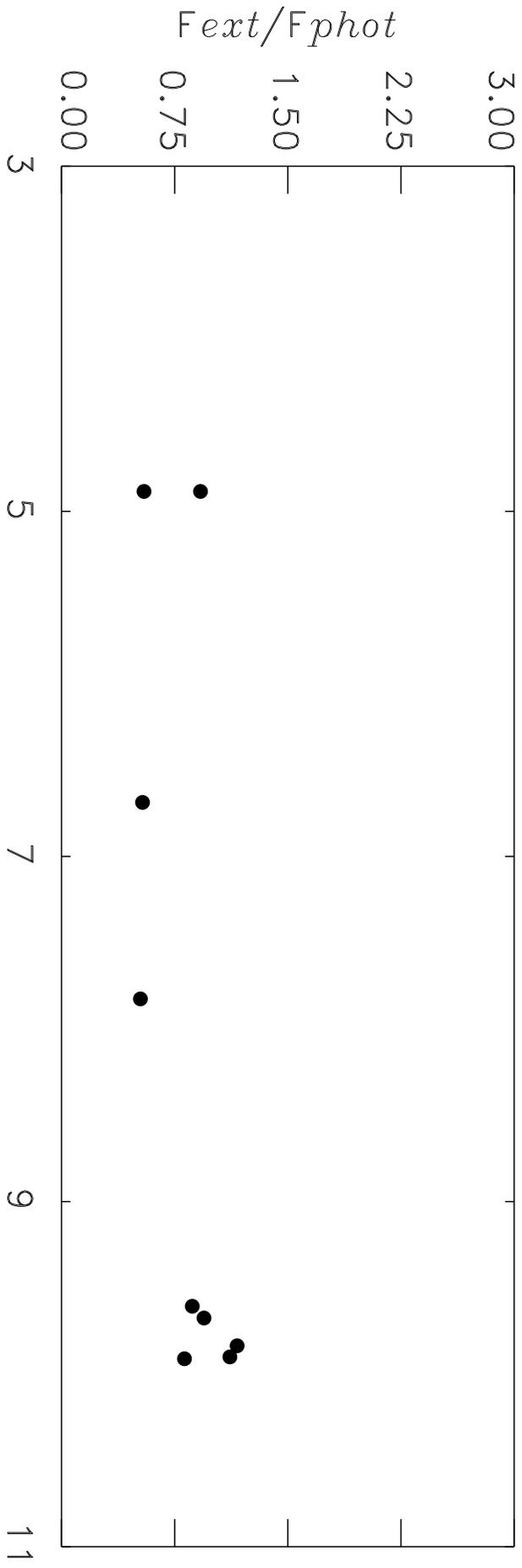
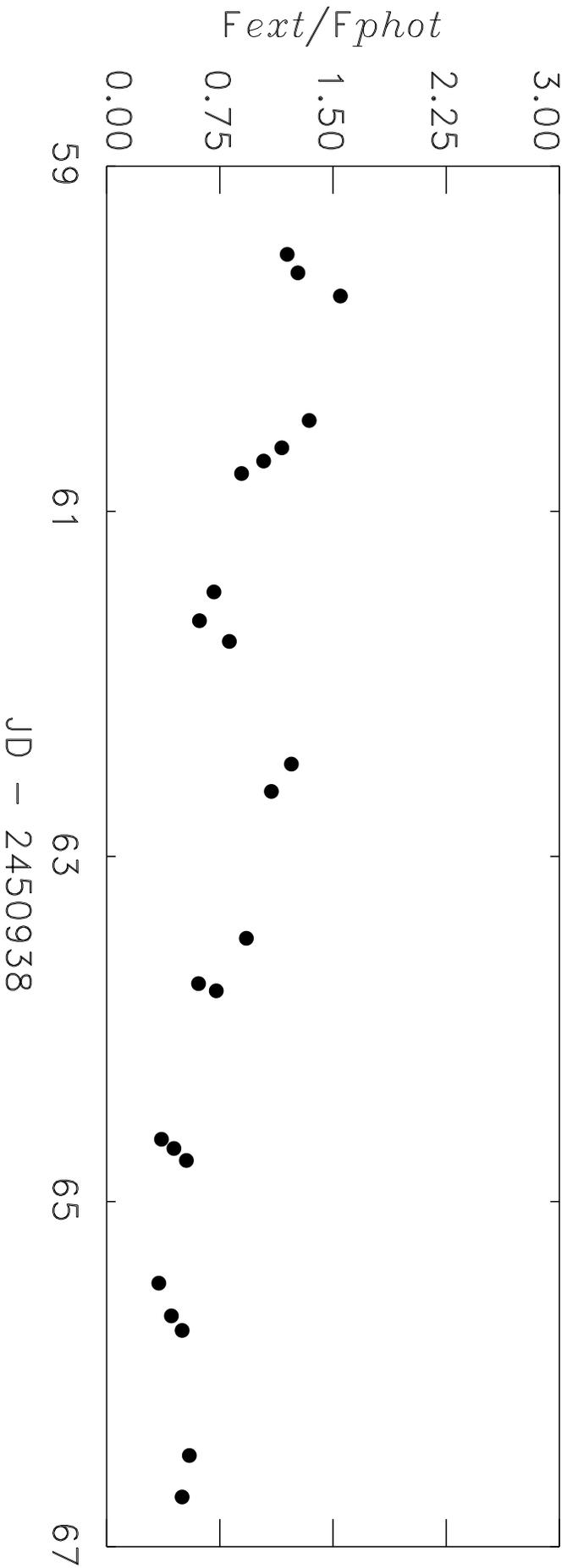
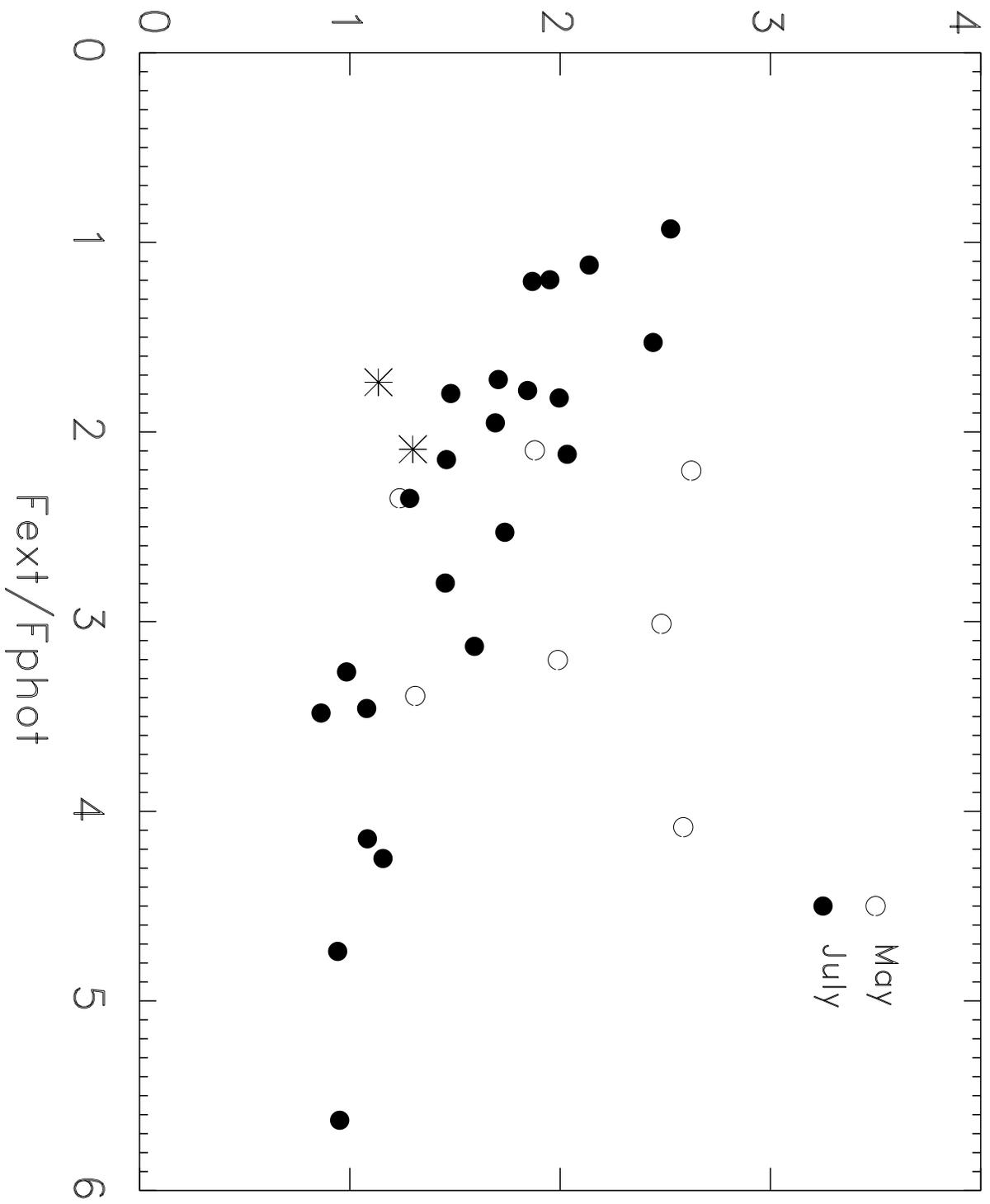
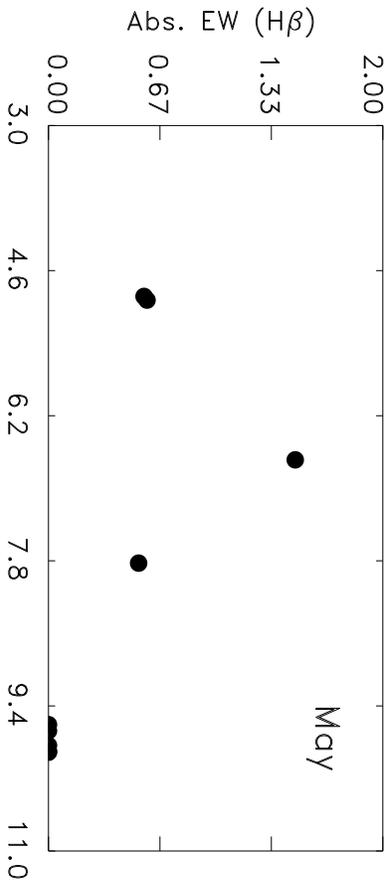
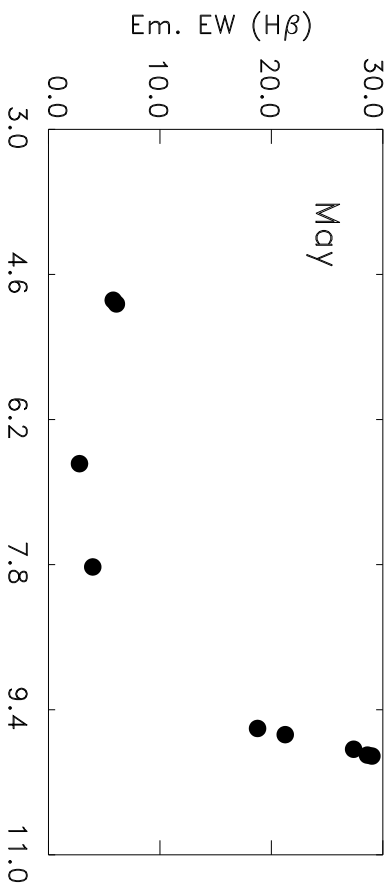
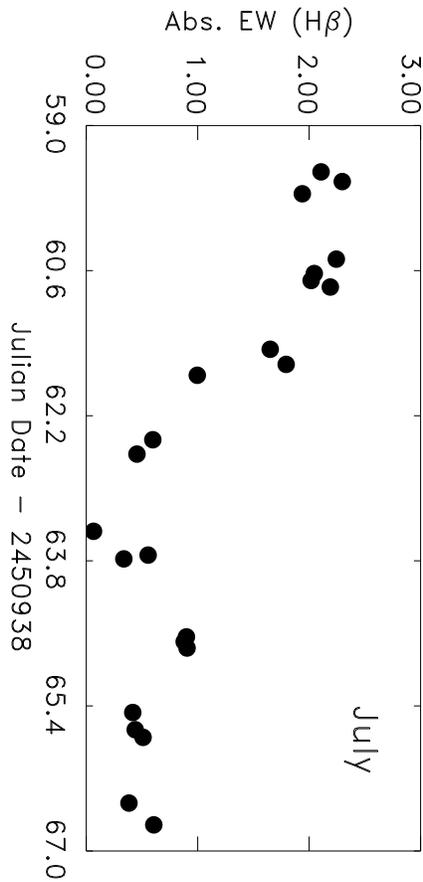
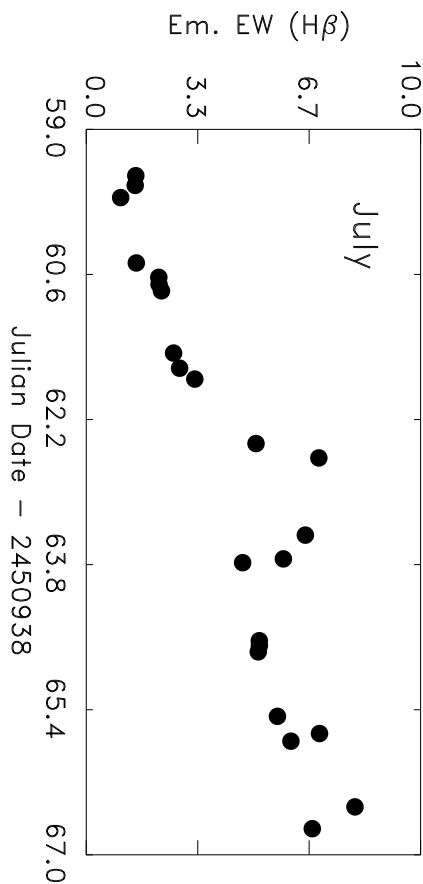
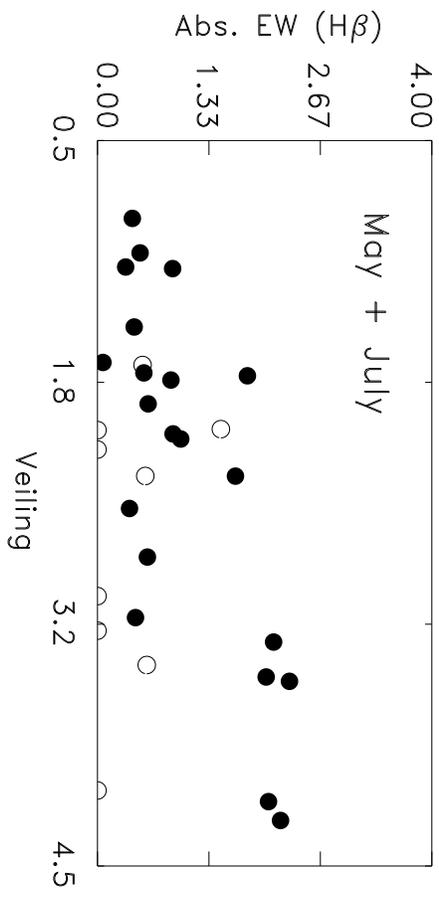
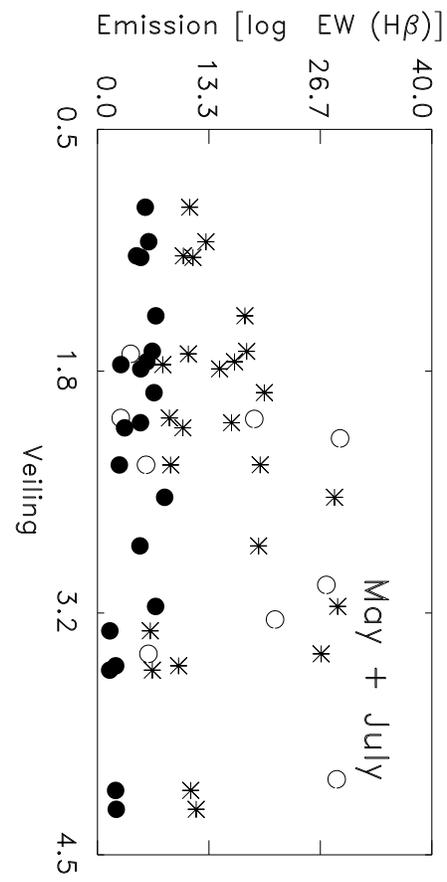
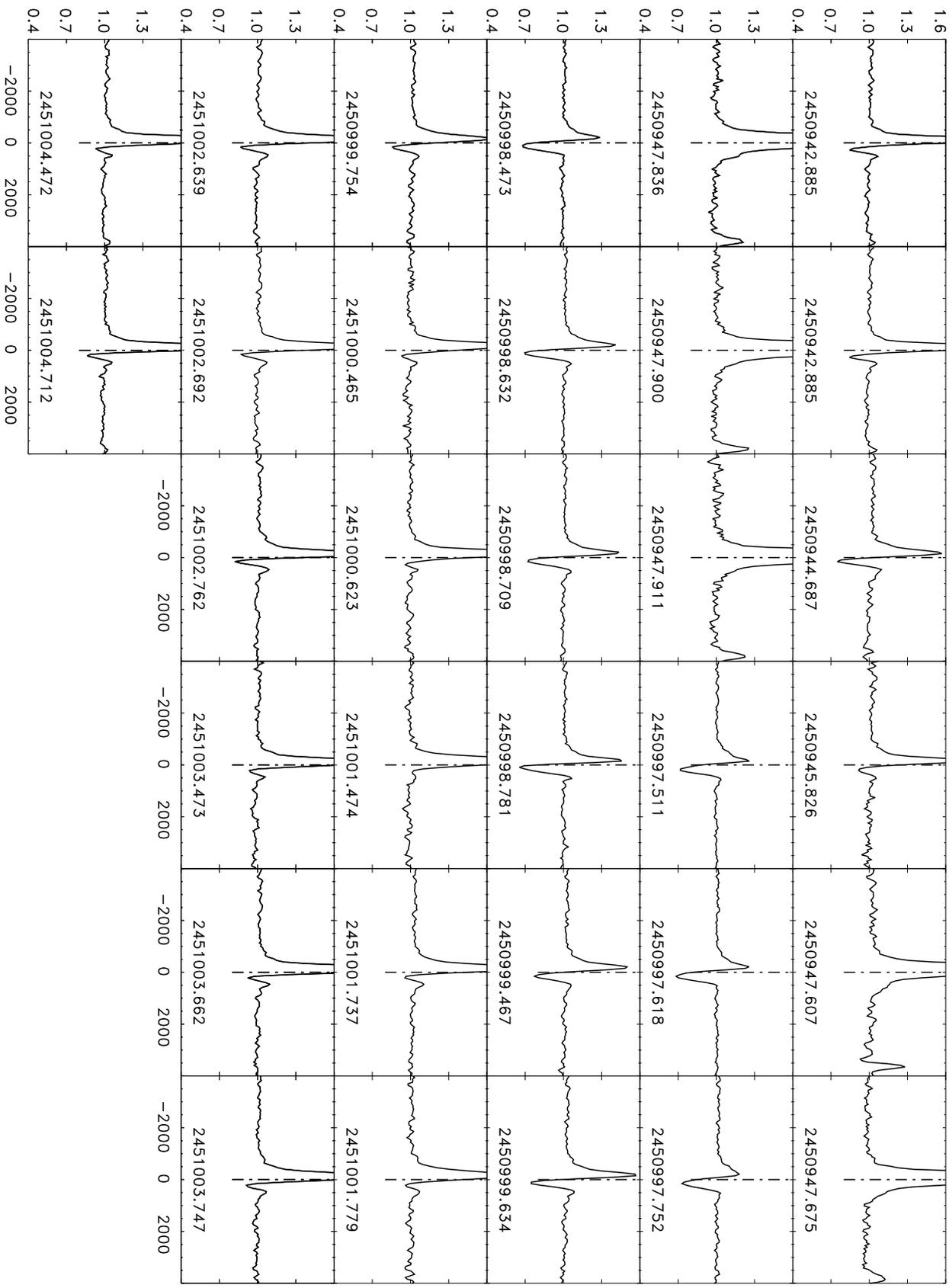
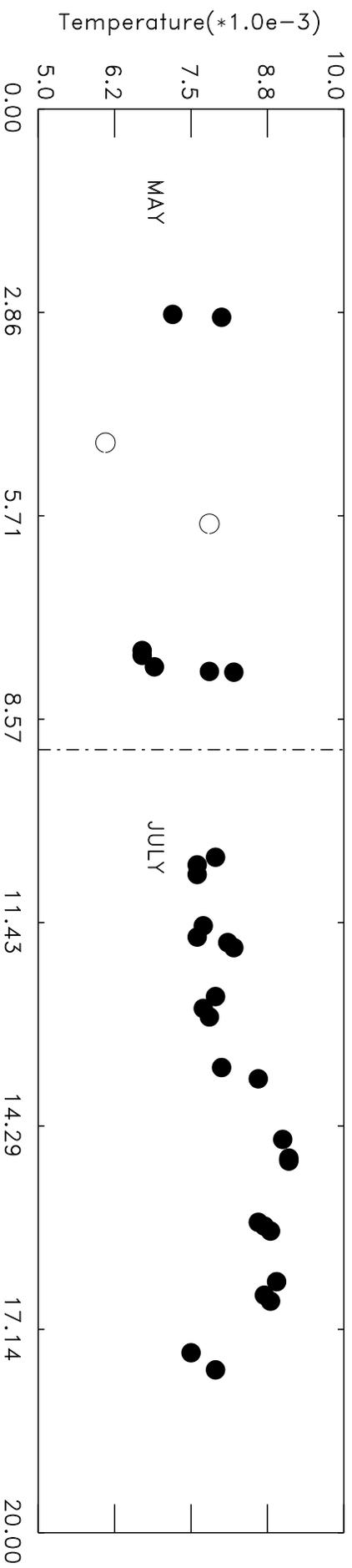
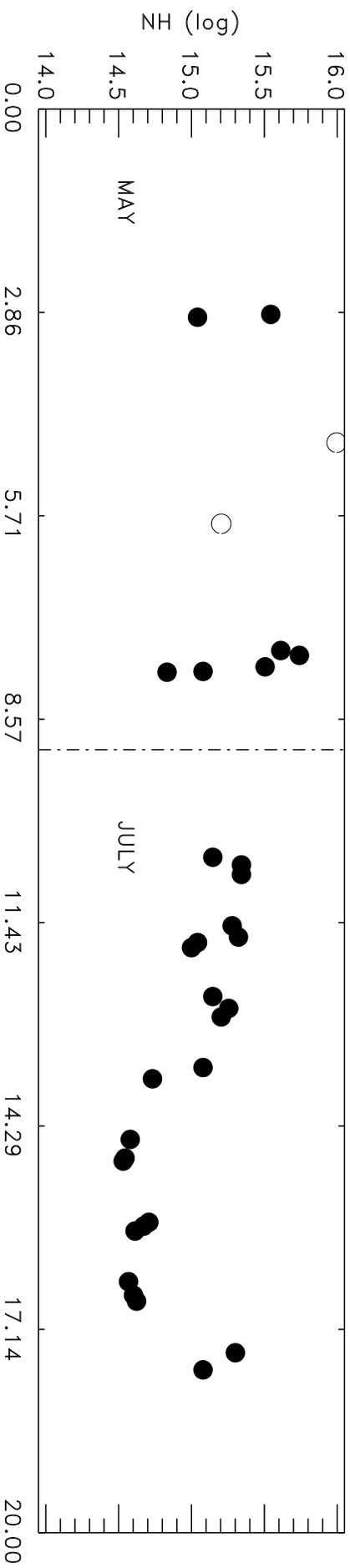
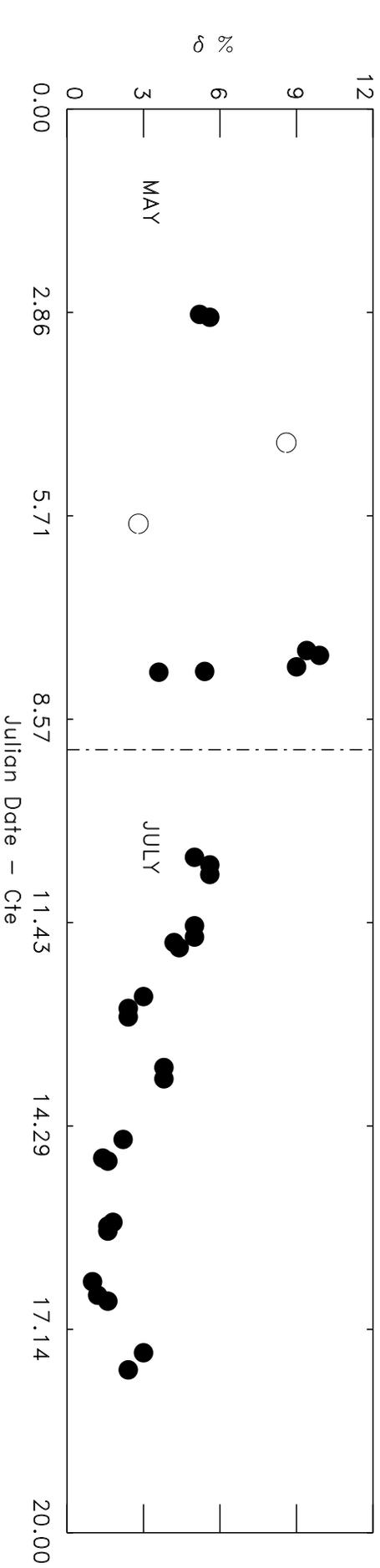
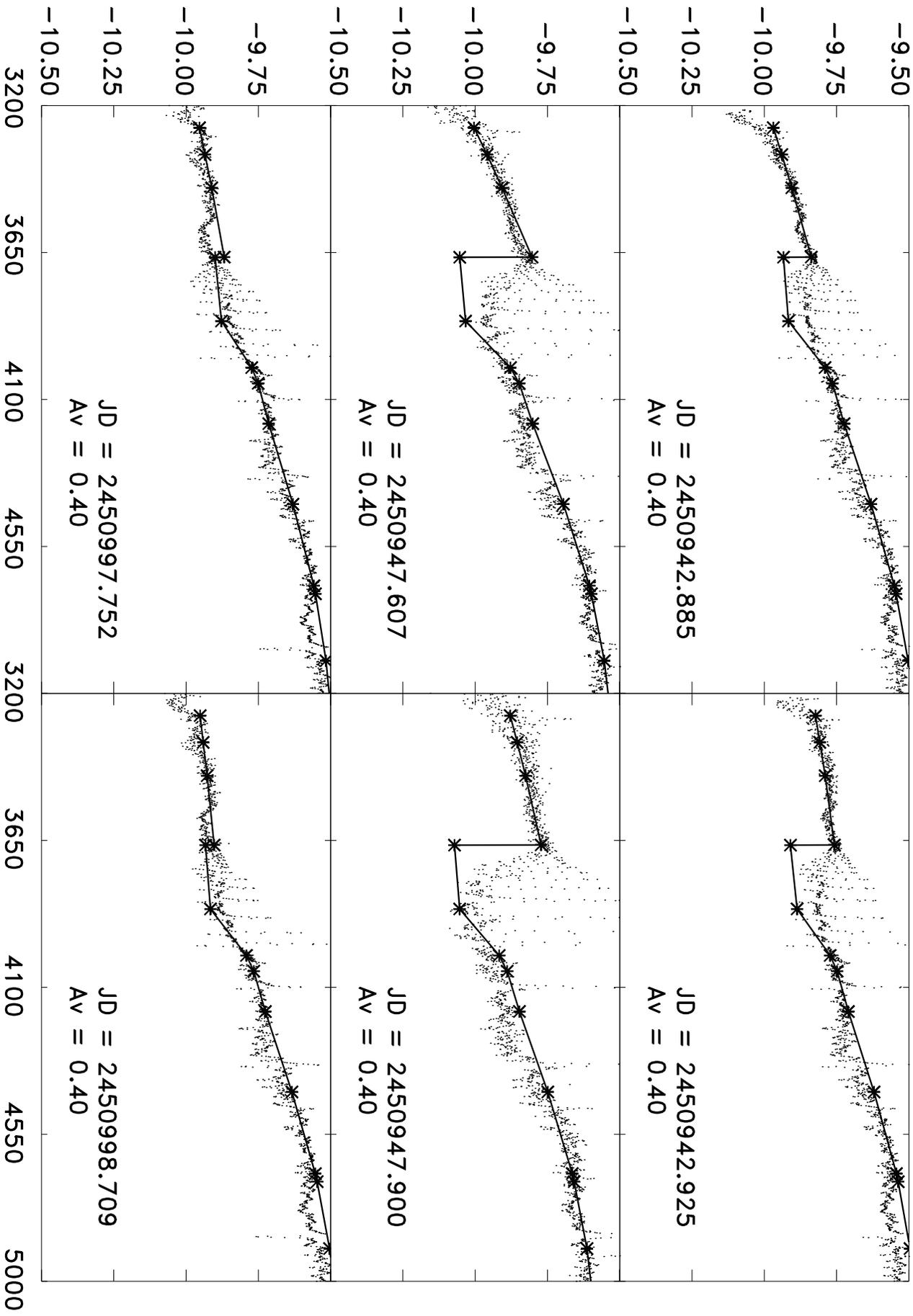
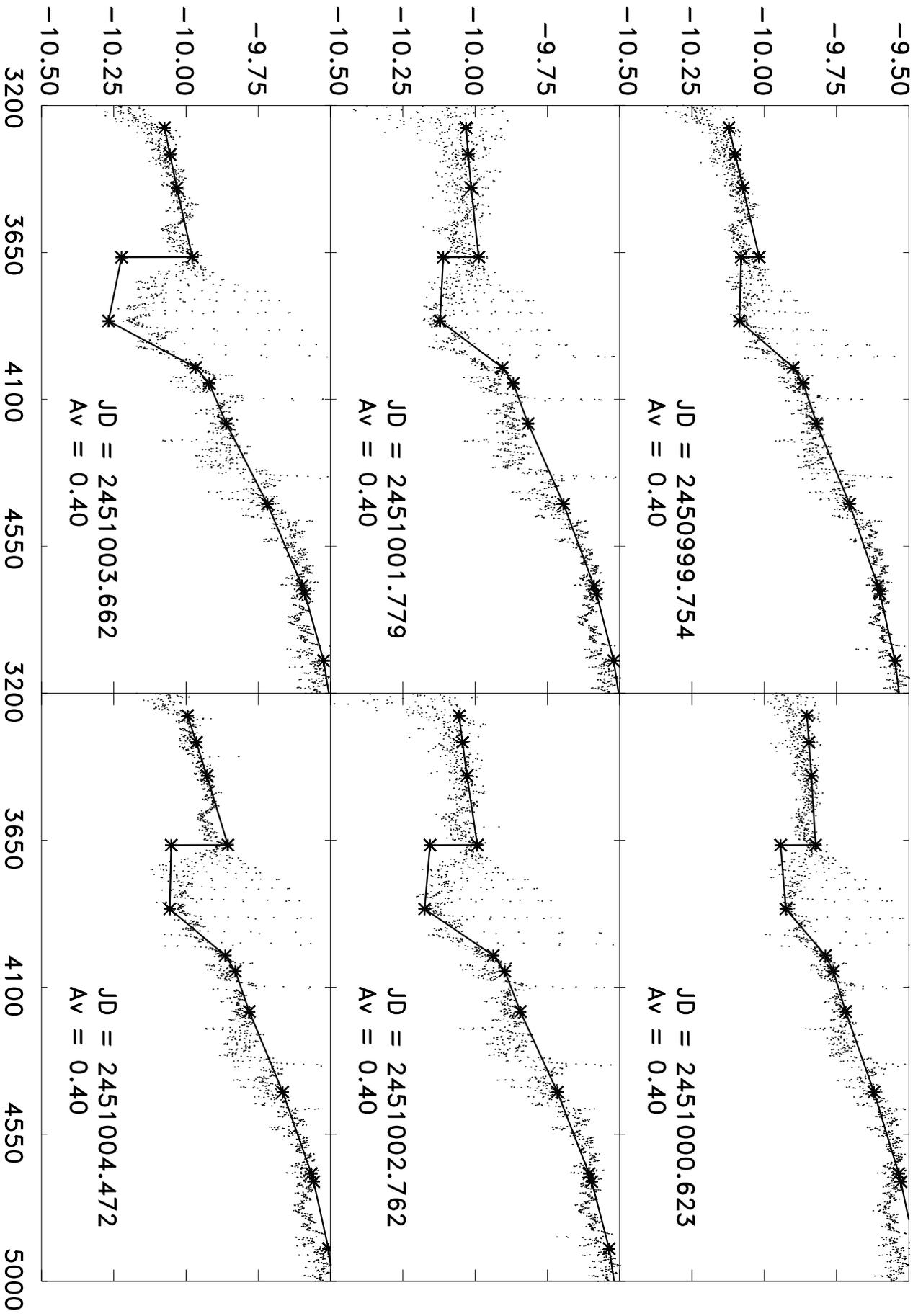
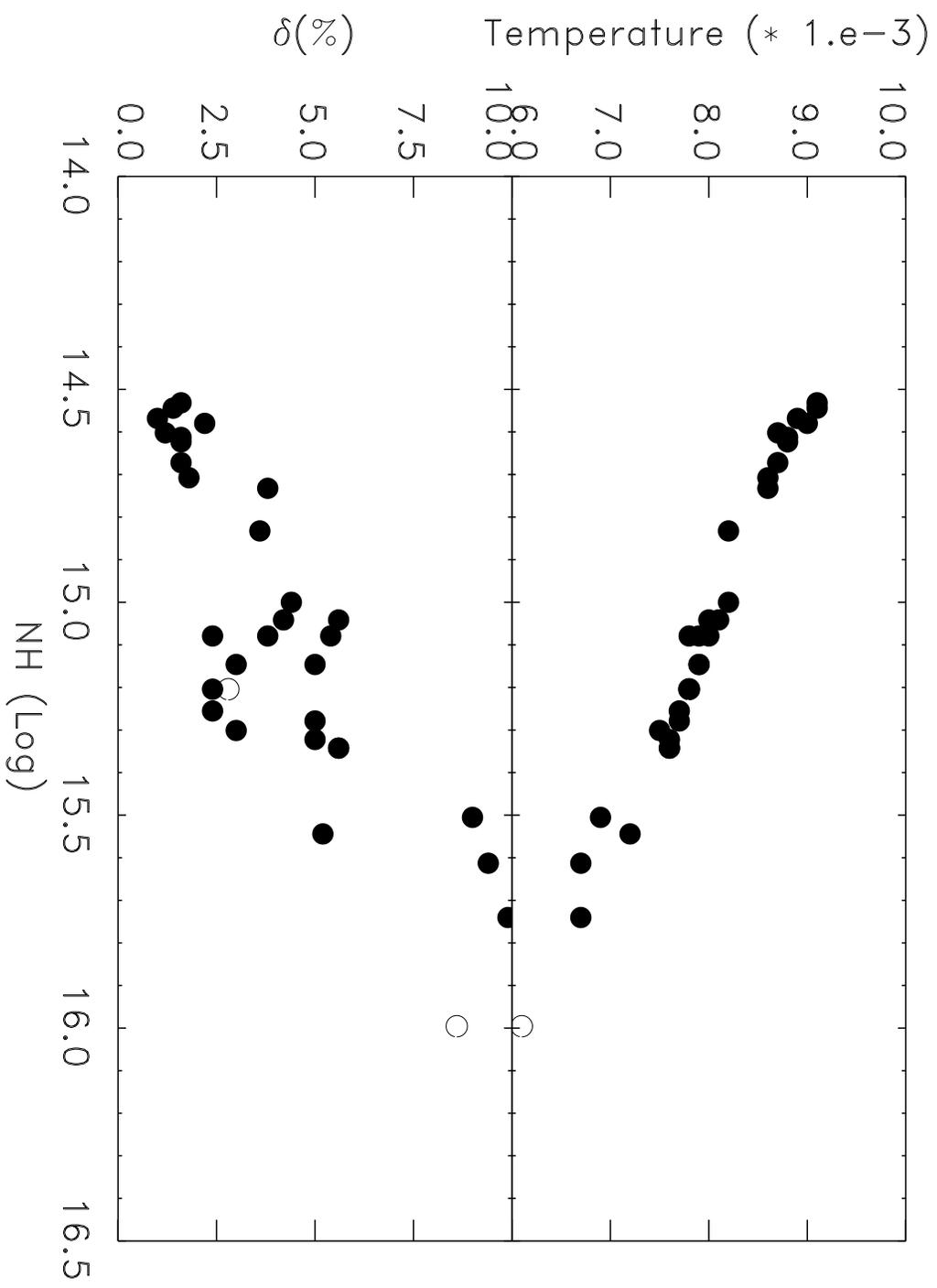
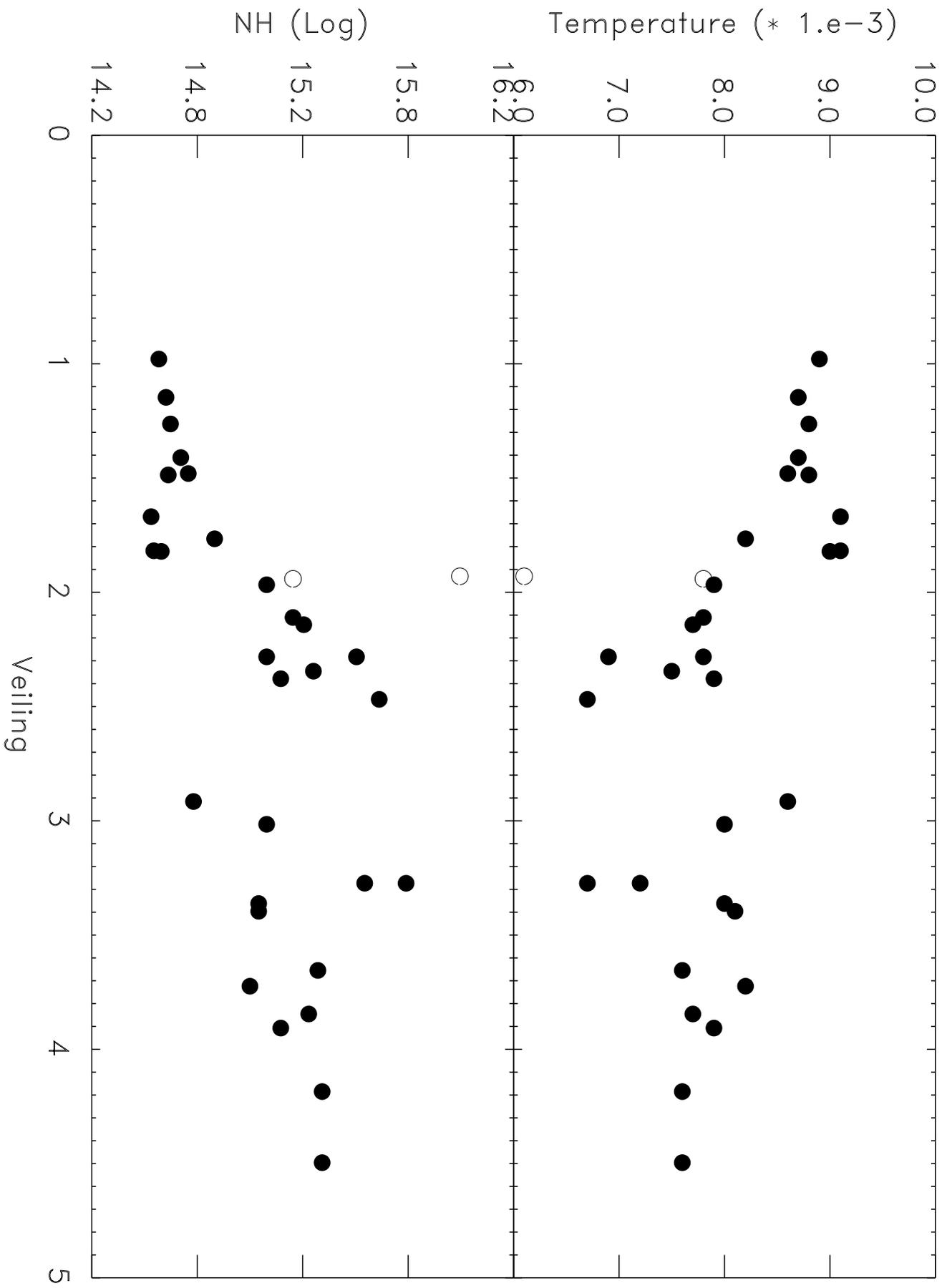
\end{document}